# A Comprehensive Review of Multimodal XR Applications, Risks, and Ethical Challenges in the Metaverse.

**Panagiotis Kourtesis**[1-4]*

[1] Department of Psychology, The American College of Greece, 15342 Athens, Greece
[2] Department of Informatics & Telecommunications, National and Kapodistrian University of Athens, 16122, Athens, Greece
[3] Department of Psychology, National and Kapodistrian University of Athens, 157 84 Athens, Greece
[4] Department of Psychology, The University of Edinburgh, Edinburgh EH8 9Y, UK
* Correspondence: pkourtesis@acg.edu

**Abstract:** This scoping review examines the broad applications, risks, and ethical challenges associated with Extended Reality (XR) technologies, including Virtual Reality (VR), Augmented Reality (AR), and Mixed Reality (MR), within the context of Metaverse. XR is revolutionizing fields such as immersive learning in education, medical and professional training, neuropsychological assessment, therapeutic interventions, arts, entertainment, retail, e-commerce, remote work, sports, architecture, urban planning, and cultural heritage preservation. The integration of multimodal technologies—haptics, eye-tracking, face- and body-tracking, and brain-computer interfaces—enhances user engagement and interactivity, playing a key role in shaping the immersive experiences in the Metaverse. However, XR's expansion raises serious concerns, including data privacy risks, cybersecurity vulnerabilities, cybersickness, addiction, dissociation, harassment, bullying, and misinformation. These psychological, social, and security challenges are further complicated by intense advertising, manipulation of public opinion, and social inequality, which could disproportionately affect vulnerable individuals and social groups. This review emphasizes the urgent need for robust ethical frameworks and regulatory guidelines to address these risks while promoting equitable access, privacy, autonomy, and mental well-being. As XR technologies increasingly integrate with artificial intelligence, responsible governance is essential to ensure the safe and beneficial development of the Metaverse and the broader application of XR in enhancing human development.

## 1.    Introduction

Extended Reality (XR), which encompasses Virtual Reality (VR), Augmented Reality (AR), and Mixed Reality (MR), represents a transformative leap in how humans interact with digital environments. These immersive technologies extend users' cognitive and physical capacities into virtual spaces, fundamentally reshaping experiences across a wide range of fields, including education, healthcare, entertainment, and professional training [1], [2]. XR enables users to engage in multisensory and interactive experiences, a subject that has been extensively researched in recent years, as advances in hardware and software have enhanced the quality, accessibility, and applications of XR technologies [3], [4]. This growing interest has led to substantial empirical findings that illuminate both the opportunities and risks associated with XR across different sectors.

One of the most significant developments in XR is its role in the emerging concept of the Metaverse, a persistent, shared digital universe where virtual and physical realities converge [5]. The Metaverse aims to offer users a seamless integration of digital and physical experiences, fostering new modes of social interaction, collaboration, and productivity. XR technologies, by facilitating deep immersion in virtual environments, serve as a foundational interface for the Metaverse. Through XR, users can transition between real and virtual spaces, participate in virtual meetings, interact with digital content, and experience simulations that mirror real-world environments, thus driving new ways of working, learning, and socializing [6], [7].

XR's importance for the Metaverse lies in its ability to offer embodied experiences, where users interact with digital content as if it were part of the physical world. This capacity for real-time, spatial interaction is essential for the Metaverse's vision of an immersive, three-dimensional space that supports not only entertainment but also professional, educational, and social activities [8]. For example, XR-based platforms in professional training and education have already demonstrated their potential in simulating complex tasks, such as medical procedures and architectural designs, within the Metaverse [9], [10]. Furthermore, XR's ability to foster social presence, where users feel physically situated within virtual environments, enhances the quality of interactions within the Metaverse, allowing for more natural communication and collaboration [4].

*1.1. Relevant Works on XR technologies and the Metaverse*

The field of XR has experienced significant growth over the past decade, with a sharp increase in research publications in the last five years. Studies have explored XR's applications in diverse domains, including healthcare, education, entertainment, and professional training. A notable example is the study by [10], which provides a comprehensive analysis of how XR technologies have been integrated into vocational training within the construction industry. This paper highlights XR's role in improving safety, spatial awareness, and operational efficiency by enabling workers to practice complex tasks in simulated environments without the risks associated with real-world training. The study underscores XR's ability to offer hands-on, experiential learning that replicates real-world scenarios, a characteristic that is also highly relevant in sectors such as healthcare and education [9], [11].

In the domain of education, Metaverse was seen as an immersive learning environment [5], a survey which offers a detailed exploration of how XR technologies, specifically within the context of the Metaverse, are reshaping educational paradigms. This review emphasizes the potential of XR to create immersive and interactive learning environments that allow students to engage with content in a more embodied and experiential manner. As [5] points out that the Metaverse, powered by XR, enables learners to access simulated real-world environments, making it easier to practice skills that are difficult to learn through traditional methods. For instance, medical students can perform surgeries in virtual environments, and engineering students can interact with complex machinery without physical limitations. This review also discusses the pedagogical benefits of XR, such as enhancing motivation and engagement through gamified learning experiences and collaborative virtual classrooms, where learners can interact in real-time, irrespective of geographical barriers. This focus on embodied learning and experiential education differentiates XR from other digital learning tools, reinforcing its unique role in modern pedagogy [5], [11].

In contrast, literature reviews such as [12] have focused on the ethical and privacy concerns associated with XR. As XR platforms rely heavily on data collection, including biometric, behavioral, and cognitive data, there are significant risks regarding user privacy and data security. This paper highlights the privacy challenges posed by XR's immersive environments, where vast amounts of sensitive data are collected to enhance user experiences. These concerns are exacerbated in the context of the Metaverse, where the integration of AI with XR could lead to more pervasive data collection and surveillance, raising questions about consent, data ownership, and potential misuse [3], [12]. The review of [12] emphasize that as users spend more time in immersive environments like the Metaverse, the aggregation of personal data becomes more intricate and potentially intrusive, leading to a need for stricter data governance frameworks to protect users' privacy and autonomy.

In addition to privacy issues, other literature has explored the psychological and social impacts of XR. For instance, [13] discusses how XR technologies can significantly influence users' emotions, cognition, and behavior. His research outlines the potential for both positive outcomes, such as enhanced learning and empathy building, and negative consequences, including addiction, dissociation, and psychological manipulation. These concerns are particularly relevant in the context of hyper-targeted advertising and misinformation within immersive environments like the Metaverse, where users are more susceptible to emotional and cognitive influence due to the immersive nature of the technology [2], [13]. The work of [13] also highlights the potential of XR to affect users' mental health, particularly in highly immersive scenarios where the distinction between reality and virtuality may become blurred, raising questions about long-term psychological impacts.

Further empirical research has explored the application of XR in healthcare, with studies examining its use for medical training, patient therapy, and rehabilitation. The study of [4] provides a detailed review of how XR has been used in clinical settings, particularly in therapies for psychological disorders such as PTSD, anxiety, and phobias. The immersive nature of XR allows for controlled exposure to triggers in a safe environment, making it a valuable tool for behavioral therapy. Similarly, in rehabilitation, XR technologies have been used to simulate real-world activities for patients recovering from injuries, helping them regain physical mobility and cognitive functions through virtual practice [14]. In addition, XR is increasingly used for pain management and distraction therapy, offering patients a non-pharmacological way to manage chronic pain and discomfort through immersive virtual environments [15].

While these studies demonstrate the wide-ranging applications of XR, the current review distinguishes itself by offering a broader synthesis of empirical data across multiple sectors, including healthcare, education, entertainment, and professional training. Furthermore, this review discusses the multimodal interactions and user's data, which appears an integral part of the interaction and user experience in the Metaverse. In addition to discussing XR's practical applications, this review delves deeper into the psychological and ethical challenges posed by these technologies, such as the risks of user dissociation, addiction, harassment, bullying, cybersecurity and the erosion of autonomy in immersive environments among others. These issues, while acknowledged in the literature, have not yet been thoroughly explored, particularly in the context of the Metaverse, where the convergence of virtual and physical realities could amplify these risks [12], [13]. Moreover, as XR technologies become more integral to daily life and professional activities, the need for comprehensive ethical guidelines and regulatory frameworks becomes more urgent, particularly to safeguard user privacy, psychological well-being, and autonomy.

*1.2.   XR's Importance for the Metaverse*

XR technologies are foundational to the development of the Metaverse, a digital universe where virtual and physical realities merge to create a persistent, immersive environment. The immersive qualities of XR—its ability to provide multisensory, embodied experiences—are critical for realizing the Metaverse's potential as a platform for work, education, entertainment, and social interaction [5]. As the Metaverse aims to integrate digital interaction into daily life, XR's role in enabling seamless transitions between real and virtual environments is essential. For instance, VR allows users to inhabit fully digital spaces, while AR overlays digital information onto the physical world, and MR blends elements of both, enabling interaction with digital and physical objects simultaneously [1], [16].

Moreover, XR enhances social presence within the Metaverse, allowing for more natural, intuitive interactions that mimic real-world dynamics. This is particularly important in professional settings, where collaboration, communication, and teamwork are essential. XR's ability to support embodied experiences—where users' actions in the virtual world reflect their physical movements—makes it a crucial technology for the Metaverse's development into a platform that supports complex tasks and interactions, such as virtual workspaces, educational simulations, and entertainment venues [6], [7].

*1.3.   This Review Aims and Structure*

The growing body of literature on XR underscores its transformative potential across various sectors. However, as XR becomes integral to the development of the Metaverse, ethical concerns surrounding privacy, data security, and psychological well-being must be addressed. This review aims to provide a comprehensive analysis of XR technologies, highlighting both the opportunities and risks they present, and offering insights into how these technologies can be deployed responsibly in the evolving digital landscape. Specifically, this review is structured as follows:

Section 2, "Multimodal Interaction Across the Virtual Continuum," delves into the immersive capabilities of XR technologies, examining how different forms of XR enhance user interaction. The section begins by discussing the Reality-Virtuality Continuum, exploring the differences between VR, which creates fully immersive digital environments, and AR and MR, which blend virtual elements with the real world. Following this, the section details key interaction methods that XR technologies enable, including Eye-Tracking for visual engagement, Hand-Tracking and Gesture Control for natural interaction with virtual objects, Haptic Feedback for adding the sense of touch, Facial Tracking for emotional and social presence, Galvanic Skin Response (GSR) for tracking emotional arousal, and Electroencephalogram (EEG) for monitoring cognitive load and brain activity. Each of these modalities is critical in enhancing the immersive experience and improving the effectiveness of interactions in XR environments.

Section 3, "XR Applications: Expanding Multimodal Interactions Across Domains," explores the vast array of applications of XR technologies across multiple sectors. It begins with the Clinical Applications of XR in areas such as therapy, neuropsychological assessment, rehabilitation, and pain management. Next, the section examines Educational Applications, highlighting XR's role in enhancing STEM education, historical simulations, language learning, and special education. It also covers the application of XR in Professional Training and Skill Development, including fields like medical training, emergency response, manufacturing, and aviation. The section continues by addressing XR's role in Arts and Entertainment, Public Health and Safety Training, Retail and E-Commerce, Architecture and Urban Planning, Sports Training, Remote Work, and Museums and Cultural Heritage Preservation. Each application demonstrates the transformative potential of XR in real-world contexts, enabling more immersive, interactive, and effective experiences.

Section 4, "Potential Risks and Ethical Challenges of XR and the Metaverse," addresses the various risks associated with XR technologies. The section begins with a discussion of Cybersickness, including its prevalence, contributing factors, and potential technological solutions. Next, it covers Mental Health and Behavioral Risks, focusing on the dangers of addiction and dissociation in XR environments. It also examines the increasing risks of Cyber Harassment and Cyberbullying, emphasizing the unique challenges of moderating user behavior in immersive virtual spaces. The section then explores Data Privacy and Security Risks, highlighting the ethical concerns associated with the vast amounts of biometric, behavioral, and emotional data collected in XR environments. This is followed by a discussion of Intense Advertising and Commercial Exploitation, which explores how hyper-targeted and manipulative advertising practices may exploit users' cognitive and emotional states. The section also addresses the manipulation of public opinion through Misinformation, Deepfakes, and AI-generated content, alongside the Physical Health Concerns arising from prolonged XR use, such as eye strain, musculoskeletal issues, and sedentary lifestyles. Additionally, the section touches on the Digital Divide and Inequality, stressing how economic and infrastructural barriers may prevent equal access to XR technologies.

Finally, Section 5, "General Discussion," synthesizes the insights gained from the preceding sections, providing an integrated reflection on the applications, risks, and broader implications of XR technologies. The discussion covers the societal and technological impacts of XR, including the Digital Divide and the need for equitable access to these transformative technologies. This section also explores future integration with Artificial Intelligence (AI), considering how AI could enhance personalization and real-time adaptation in virtual environments. The discussion underscores the importance of maintaining a balanced approach to the development and implementation of XR technologies to ensure their benefits are realized while addressing key ethical and psychological concerns.

## 2. Multimodal Interaction Across the Virtual Continuum

### 2.1. Reality–Virtuality Continuum

The Reality-Virtuality Continuum, introduced by Milgram and Kishino (1994) [1], illustrates the interaction between physical and virtual realities, ranging from entirely physical environments to fully virtual ones. Augmented Reality (AR) and Mixed Reality (MR) occupy intermediate positions, blending real and digital worlds. XR technologies—covering VR, AR, and MR—offer various degrees of immersion, allowing users to engage with digital environments in increasingly integrated ways. As immersive technologies evolve, the continuum helps explain how XR augments our physical interactions and cognitive processes, like memory and spatial reasoning, positioning XR at the core of the emerging Metaverse, where physical and virtual experiences converge.

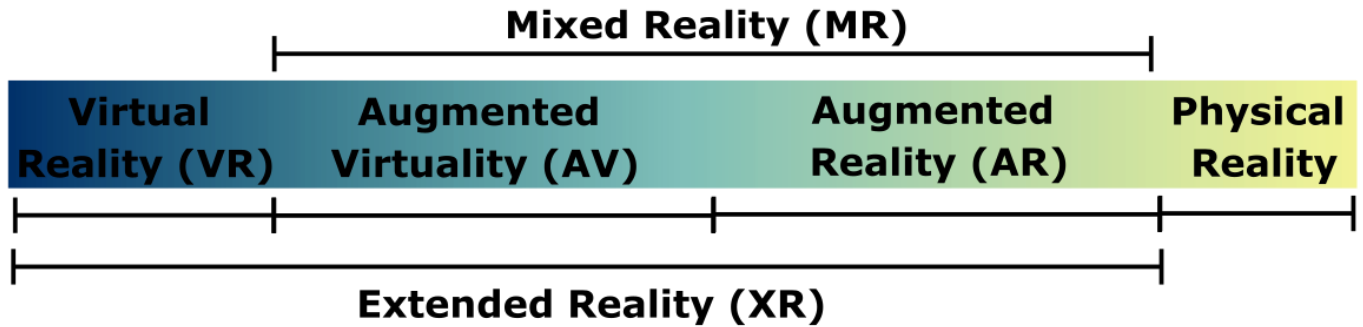

Figure 1. The Virtual Continuum: From the Physical to the Virtual, with AR and MR blending elements of both worlds.

#### 2.1.1. Virtual Reality (VR) and Cognitive Immersion

At one end of the continuum, VR fully immerses users in a synthetic environment, replacing physical surroundings. Using head-mounted displays (HMDs) and controllers, VR offers strong sensory feedback through visuals, sound, and haptics, promoting a deep sense of presence [17]. VR enhances applications like education, healthcare, and therapy by creating immersive spaces for learning and treatment, reducing cognitive load through virtual support for memory and spatial reasoning tasks [18]. Multimodal technologies such as eye- and hand-tracking further improve interaction, making VR highly engaging and responsive.

#### 2.1.2. Augmented Reality (AR) and Real-World Integration

At the opposite end, AR overlays digital content onto the physical world, enriching real-world experiences without replacing them. Accessed via devices like smartphones or AR glasses, AR adds layers of information, making it ideal for applications in healthcare, education, and industrial training. AR helps visualize real-time data during surgeries or educational exercises, offering interactive learning experiences. Multimodal features like gesture control and eye-tracking enhance its usability, enabling intuitive interactions between physical and digital elements.

#### 2.1.3. Mixed Reality (MR) and Hybrid Interaction

MR sits between VR and AR, merging real and virtual elements for simultaneous interaction. MR allows users to manipulate both real and virtual objects in real-time, benefiting fields like architecture and healthcare. It enables architects to visualize structures in real-world locations or medical professionals to practice complex procedures in hybrid environments. MR's use of hand-tracking, haptics, and body-tracking enhances the tactile and immersive experience, blending physical and virtual elements seamlessly for practical, interactive applications.

*2.2.  Multimodal Interaction in XR: Enhancing Immersion and Interaction*

In XR environments, the integration of multiple sensory modalities—such as vision, touch, sound, and physiological signals—transforms how users interact with and experience virtual worlds. Multimodal interaction not only enhances immersion but also allows for more intuitive, adaptive, and cognitively enriching experiences. As XR technologies continue to evolve, these modalities work in synergy to create virtual environments that respond to the user's actions, emotions, and cognitive states, making XR one of the most interactive and engaging platforms for training, entertainment, therapy, and more (see Table 1).

Table 1. Multimodal Systems in Extended Reality

| Modality | Primary Function | Applications | Example |
| --- | --- | --- | --- |
| Eye-Tracking | Tracks visual attention and cognitive load | Education, professional training, gaming, healthcare | Adapting content based on gaze in training simulations, adjusting difficulty based on cognitive load |
| Facial Tracking | Captures and replicates real-time facial expressions to enhance emotional and social presence | Virtual workspaces, social hangouts, gaming, therapy | Avatars mirroring users' facial expressions for more natural interaction |
| Hand/Finger Tracking | Allows intuitive manipulation of virtual objects using hand and finger gestures | Professional training, collaborative work, gaming, creative industries | Manipulating virtual objects in real time during design sessions or medical simulations |
| Full-Body Tracking | Tracks full-body movements for realistic avatar replication and physical engagement | Fitness, physical therapy, social interaction, sports training | Performing physical exercises in a virtual fitness program, analyzing posture in physical therapy |
| Haptic Feedback | Provides tactile sensations to simulate the feeling of virtual objects and environments | Training simulations, gaming, therapy, creative industries | Feeling resistance when handling virtual tools in mechanical training or receiving tactile feedback in creative tasks |
| Galvanic Skin Response (GSR) | Monitors emotional states through skin conductance linked to arousal levels | Therapy, stress management, emotional engagement | Adjusting virtual environments in real time based on stress or relaxation levels detected via GSR |
| Heart Rate Monitoring | Tracks physiological engagement through heart rate changes | Virtual fitness, stress management, emotional engagement | Adjusting workout intensity in virtual fitness programs based on heart rate feedback |
| EEG (Electroencephalogram) | Measures brain activity and cognitive states for brain-computer interaction (BCI) | Accessibility, cognitive control, education, professional training | Allowing users to control virtual objects with brain activity or adjusting task difficulty based on cognitive load data |

2.2.1.   Eye-Tracking: Enhancing Visual Engagement and Cognitive Interaction

Eye-tracking technology has emerged as one of the most transformative modalities in XR environments, offering a dynamic way for systems to monitor and respond to where users focus their visual attention. By capturing real-time data on a user's gaze, eye-tracking systems can enhance both the responsiveness and interactivity of virtual environments. This modality is particularly powerful because it allows the virtual environment to adapt dynamically based on user behavior, reducing cognitive load and enhancing user engagement [19].

For example, in educational XR applications, eye-tracking can highlight key elements of the virtual environment as users focus on specific objects, helping them to better engage with learning materials. If a student is looking at a complex molecular structure, for instance, the system can offer additional information or zoom in to provide a clearer view, making the learning experience more personalized and efficient. In professional training, eye-tracking can be used to monitor attention and ensure that users are focusing on critical information during simulations [20], [21], [22].

In addition to interaction and engagement, eye-tracking provides valuable cognitive insights by monitoring gaze patterns such as fixation durations (how long a user focuses on an object) and saccades (rapid eye movements between points of focus). These metrics allow systems to assess cognitive load—whether a task is too easy or too difficult for the user—and adjust the environment accordingly. For instance, if the system detects prolonged fixation on a complex object, indicating that the user is experiencing cognitive strain, it can reduce visual distractions or simplify the environment to aid comprehension [23], [24]. Conversely, if cognitive load is low, the system may introduce more challenging tasks or information to maintain engagement.

Eye-tracking also enables gaze-based interaction within XR environments, allowing users to interact with virtual objects simply by looking at them. This intuitive form of interaction reduces the reliance on traditional input devices, such as controllers, making the experience more natural and immersive. For instance, users can select virtual tools, navigate menus, or trigger animations based on their gaze, creating a seamless interface between the user and the digital world [25]. Gaze-based interaction is particularly beneficial in fields like healthcare, where medical professionals can quickly access digital patient files or instructional overlays by simply focusing on specific parts of the interface, improving both efficiency and accuracy [26].

2.2.2.   Hand-Tracking and Gesture Control: Natural Interaction with Virtual Objects

Hand-tracking and gesture control technologies have become critical components in enhancing the realism of XR experiences, allowing users to interact with virtual objects as they would in the real world. These technologies capture fine movements of the user's hands and fingers, enabling natural manipulation of digital elements. By eliminating the need for external controllers, hand-tracking makes the interaction more fluid and intuitive, reducing barriers between the user and the virtual environment.

In education and professional training, hand-tracking enables users to engage in hands-on activities within a virtual setting. For example, medical students can use hand gestures to practice complex surgical procedures in a virtual operating room, manipulating instruments with precision and receiving real-time feedback. In engineering and design fields, hand-tracking allows professionals to interact with virtual prototypes, making adjustments to 3D models with the same dexterity they would use on physical objects, fostering creativity and iterative problem-solving [27, p. 20].

Hand-tracking also enhances collaboration in XR environments, where multiple users can interact with the same virtual objects in real time. In a virtual design studio, for instance, architects from different locations can simultaneously manipulate a building model, making adjustments collaboratively and offering real-time feedback. This kind of shared, interactive engagement fosters teamwork and accelerates the design process by allowing users to "handle" virtual objects with precision [28], [29]. This is particularly valuable in industries like product design, where rapid iterations and adjustments are essential to the creative workflow.

In entertainment and gaming, hand-tracking opens new avenues for immersive gameplay. Users can wield weapons, cast spells, or manipulate virtual objects using their hands, making the gameplay more engaging and personal. For example, in a fantasy role-playing game, players can cast magic spells by making specific hand gestures, while in action games, they can block, punch, or interact with enemies using natural movements. This form of physical engagement increases the player's sense of presence in the game world and creates more dynamic, personalized experiences [30].

2.2.3.   Haptic Feedback: Adding the Sense of Touch to XR

While visual and auditory stimuli have long been central to XR environments, the addition of haptic feedback—the ability to simulate the sense of touch—has taken immersion to the next level. Haptic feedback provides users with tactile sensations, such as vibrations, resistance, or pressure, that mimic the experience of interacting with physical objects. This sensory modality allows users to "feel" virtual objects, making interactions in XR environments more tangible and realistic [31].

In training simulations, haptic feedback is particularly valuable. For example, in medical training, haptic gloves or controllers can simulate the resistance of human tissue, allowing users to practice delicate surgical procedures with greater precision. This tactile feedback is crucial for developing muscle memory and fine motor skills, as it provides a realistic sense of interaction that goes beyond visual or auditory cues [32], [33], [34], [35]. Similarly, in mechanical training, haptic feedback enables users to feel the weight and texture of virtual tools, making it easier to perform tasks such as assembling machinery or repairing engines.

Haptic devices are also used in gaming and entertainment to increase immersion. In virtual reality games, for instance, users might feel the recoil of a weapon or the impact of a collision, heightening the emotional and physical engagement with the virtual world. The sense of touch adds another layer to the player's experience, making it more immersive and thrilling. In virtual art studios, haptic gloves allow artists to "feel" the digital clay or paintbrush, providing a more authentic creative process that closely mirrors traditional art forms [32], [36], [37].

Looking ahead, advanced haptic systems that simulate textures, temperature, and even pain are being developed, further expanding the range of tactile experiences available in XR. These advancements will not only enhance entertainment and training applications but also open up new possibilities for therapeutic interventions, such as physical rehabilitation and emotional therapy [38].

2.2.4. Facial Tracking: Enhancing Emotional and Social Presence

Facial tracking technology plays a crucial role in social and collaborative XR environments by capturing and replicating real-time facial expressions. This modality allows virtual avatars to convey subtle emotional cues such as smiles, frowns, or raised eyebrows, enhancing the authenticity of interactions and deepening emotional connections between users. By accurately reflecting users' facial movements, facial tracking brings a new level of emotional and social presence to virtual environments [39].

In virtual meetings and collaborative workspaces, facial tracking enhances non-verbal communication, which is often lost in traditional online interactions. By replicating real-time facial expressions, it bridges the gap between face-to-face communication and virtual collaboration. This has significant implications for remote work, where gestures like a nod of approval or a skeptical look can influence group dynamics and decision-making processes. In virtual social spaces, facial tracking allows for more genuine emotional expressions, making interactions feel more personal and engaging [40].

Facial tracking also supports emotionally adaptive virtual environments, where the system can detect and respond to changes in users' emotional states. For instance, in virtual therapy sessions, if facial tracking detects signs of stress or anxiety (e.g., furrowed brows or tensed muscles), the virtual environment could adjust by introducing calming stimuli such as soothing colors or ambient sounds. This capability not only makes virtual environments more emotionally responsive but also supports personalized therapeutic interventions, helping users regulate emotions in real time [41].

2.2.5. Galvanic Skin Response (GSR): Tracking Emotional Arousal

Galvanic Skin Response (GSR) is a physiological measure that monitors changes in the skin's electrical conductance, which varies based on moisture levels. The more emotionally aroused an individual is—whether from stress, excitement, or fear—the more sweat is produced, even if it is imperceptible. This variation in skin moisture affects electrical conductivity, which GSR sensors detect and use as an indicator of emotional intensity [42].

In XR environments, GSR can be a powerful tool for monitoring users' emotional states and adjusting the virtual experience in response. For example, in virtual therapy or stress management applications, GSR can help detect when users are becoming overwhelmed or anxious. The system can respond by introducing calming elements, such as dimming the lights, lowering ambient sounds, or altering the pacing of the environment to reduce stress.

GSR serves as a vital tool in tracking users' emotional arousal within XR environments. By detecting physiological changes related to stress, excitement, or relaxation, GSR allows systems to adapt dynamically, ensuring that users remain emotionally engaged without becoming overwhelmed. Its ability to respond to emotional cues makes GSR particularly valuable in applications such as therapy, gaming, and stress management, where real-time emotional feedback can personalize and enhance user experiences. Overall, GSR provides a responsive layer of emotional intelligence that helps create more immersive and user-centered XR environments.

2.2.6. Electroencephalogram (EEG): Monitoring Cognitive Load and Brain Activity

While GSR tracks emotional arousal, Electroencephalogram (EEG) technology provides deeper insights into users' cognitive states by measuring electrical activity in the brain. EEG sensors capture brainwave patterns that reflect different levels of mental engagement, focus, and fatigue, making it an essential tool for understanding and optimizing users' cognitive load in XR environments. Unlike GSR, which focuses on emotional states, EEG gives direct feedback about the user's mental performance and cognitive effort [43].

In XR-based training simulations, EEG is invaluable for ensuring that users are operating within an optimal cognitive load. If the system detects brainwave patterns associated with mental fatigue or disengagement, it can simplify tasks, slow down the pace of interactions, or reduce the complexity of the environment. This ensures that users do not become cognitively overwhelmed, allowing them to maintain focus and perform tasks efficiently. Conversely, if EEG readings indicate low cognitive load, the system may introduce more challenging tasks or increase task complexity to maintain engagement and ensure effective learning [44].

EEG also plays a pivotal role in brain-computer interfaces (BCIs), which allow users to control virtual objects or navigate digital environments using their brain activity alone. BCIs have significant implications for accessibility, particularly for individuals with physical disabilities who may not be able to use traditional input devices. By analyzing brainwave patterns associated with specific mental commands, EEG-powered BCIs can enable users to manipulate virtual objects, move through virtual worlds, or even engage in communication, offering greater autonomy and immersion [45], [46].

Overall, EEG technology provides a deeper understanding of users' cognitive processes, offering real-time feedback on mental states such as focus, engagement, and fatigue. By tracking brainwave patterns, EEG enables XR systems to adjust task difficulty and pacing, ensuring that users remain cognitively stimulated without becoming overloaded. This adaptability makes EEG particularly effective in education, training, and therapeutic contexts, where optimizing cognitive load is crucial for learning and performance. Additionally, EEG's role in BCIs opens up new possibilities for accessibility and interaction, further expanding the potential of XR technologies to augment human cognition and interaction.

## 3. XR Applications: Expanding Multimodal Interactions Across Domains

XR offers groundbreaking opportunities across a wide range of fields. By integrating multiple modalities—such as eye-tracking, hand-tracking, haptic feedback, EEG, and biometric sensors—XR enhances immersion, interaction, and user performance in virtual environments. These multimodal systems are critical in optimizing the applications of XR, making it an essential tool for clinical interventions, education, professional training, entertainment, and beyond (see Table 2). Below is an exploration of XR's transformative potential across various fields, highlighting the role of multimodal integration in each context.

**Table 2.** XR Applications with Multimodal Integration

| Domain | Key Applications | Multimodal Integration | Example |
|---|---|---|---|
| Clinical | Therapy, neuropsychological assessment, physical rehabilitation, pain management | Eye-tracking, GSR, EEG, haptic feedback | XR exposure therapy for anxiety disorders, stroke rehabilitation with motor skill tracking |
| Education | STEM education, history, language learning, special education | Hand-tracking, eye-tracking, haptic feedback, EEG | Virtual dissections in biology, historical reenactments, personalized learning for students with learning disabilities |
| Professional Training | Medical training, emergency response, manufacturing, corporate skill development | Hand-tracking, haptic feedback, EEG, GSR | XR simulations for surgical training, crisis management in disaster scenarios, leadership training |
| Arts & Entertainment | Gaming, virtual art galleries | Facial tracking, hand-tracking, haptic feedback, eye-tracking | Immersive multiplayer gaming with avatar mirroring, virtual museum experiences with tactile feedback |

| Domain | Key Applications | Multimodal Integration | Example |
| --- | --- | --- | --- |
| Public Health & Safety | Disaster preparedness, public health campaigns, workplace safety | Eye-tracking, hand-tracking, haptic feedback, full-body tracking | Virtual earthquake simulations for first responders, XR for health safety training |
| Retail & E-commerce | Virtual shopping, product customization, enhanced customer insights | Hand-tracking, eye-tracking, haptic feedback | Virtual clothing try-ons with full-body tracking, real-time eye-tracking for personalized shopping experiences |
| Architecture & Urban Planning | Virtual walkthroughs, collaborative design, safety and risk assessment | Hand-tracking, haptic feedback, eye-tracking, full-body tracking | Virtual tours of proposed buildings with real-time adjustments, AR overlays for community feedback |
| Sports Training | Skill development, mental training, performance analysis | Full-body tracking, heart rate monitoring, GSR, EEG | XR for tennis serve practice with haptic feedback, mental conditioning with GSR and EEG monitoring |
| Remote Work | Virtual meetings, remote collaboration, training, and onboarding | Eye-tracking, hand-tracking, full-body tracking, facial tracking | Interactive virtual office spaces with real-time avatar mirroring and hand-tracking |
| Museums & Cultural Heritage | Virtual museums, cultural preservation, multimodal engagement in museums | Eye-tracking, hand-tracking, haptic feedback | Digital reconstruction of endangered cultural sites, interactive museum exhibits with personalized tours |

*3.1. Clinical Applications: Therapy, Neuropsychological Assessment, and Rehabilitation*

XR technologies are making significant advancements in clinical applications, ranging from mental health therapy and neuropsychological assessment to physical rehabilitation and pain management. Through the integration of multimodal feedback systems, such as eye-tracking, facial tracking, GSR, EEG, and heart rate monitoring, XR environments can provide real-time responses to a patient's emotional and physiological states, resulting in personalized therapeutic interventions. This adaptability and personalization position XR technologies as transformative tools in clinical interventions, expanding both mental and physical health care.

3.1.1. Therapy

One of the key clinical applications of XR is VR exposure therapy, a method that has shown great promise in treating anxiety disorders, PTSD, and phobias. Patients are immersed in controlled virtual environments where they can gradually confront and manage their fears in a safe setting, allowing for more controlled and personalized treatment compared to traditional exposure therapies. A recent meta-analysis found that VR exposure therapy significantly reduces symptoms of PTSD and anxiety in a range of clinical populations [47], [48].

The integration of multimodal sensors enhances this approach by monitoring the patient's physiological responses. GSR measures stress through changes in skin conductance, while EEG provides insight into brain activity related to anxiety or relaxation. For instance, if the GSR data shows elevated stress levels during exposure to a feared stimulus (such as heights or spiders), the virtual environment can be automatically adjusted, reducing the intensity of the stimuli to encourage gradual desensitization [16], [49], [50]. This allows for a more dynamic and responsive form of therapy, where the treatment environment adapts in real time based on the patient's physiological state.

Furthermore, XR-based cognitive-behavioral therapy (CBT) is increasingly being used for the treatment of depression, addictions, and eating disorders [4], [51]. In these applications, XR's ability to simulate real-world environments allows patients to engage in therapeutic exercises that target maladaptive thoughts and behaviors in controlled, immersive settings. The combination of haptic feedback, eye-tracking, and facial recognition enables deeper emotional engagement, making therapy sessions more interactive and personalized.

### 3.1.2. Neuropsychological Assessment

XR offers enhanced tools for neuropsychological assessments, allowing for a more accurate and engaging evaluation of cognitive functions such as memory, attention, and executive functioning. Traditional assessments often rely on static paper-based tests, which can be limited in ecological validity. XR technologies provide an interactive alternative, where patients can perform tasks in immersive, dynamic environments, with their performance tracked in real time.

For instance, eye-tracking technology allows clinicians to evaluate how patients interact with stimuli during assessments, such as tracking gaze patterns to measure attention or fixations [22], [23], [52]. In tasks where memory or decision-making is involved, eye-tracking data can indicate cognitive load, revealing whether a patient is struggling to process information or focusing on irrelevant stimuli. Such detailed data allows for a more nuanced understanding of cognitive impairments, such as attention deficits, compared to traditional methods [22], [23], [52]

In addition to eye-tracking, EEG data provides insights into cognitive engagement and fatigue. Real-time EEG monitoring can detect cognitive overload, lapses in attention, or difficulty with memory retrieval, helping clinicians adjust the task's complexity during assessments [53], [54], [55]. By integrating multimodal feedback systems, clinicians can develop a more detailed and accurate cognitive profile for each patient, facilitating tailored treatment plans.

### 3.1.3. Rehabilitation

Physical rehabilitation is another area where XR is having a substantial impact. With full-body tracking and haptic feedback, patients recovering from injuries, surgeries, or neurological conditions can perform exercises in virtual environments that track and mirror their movements. For example, stroke patients can practice motor skills in a virtual environment where their movements are tracked in real-time, providing therapists with detailed data to monitor progress [56].

A 2020 review found that VR-based rehabilitation leads to significant improvements in motor function for stroke survivors, particularly when combined with multisensory feedback [55], [57], [58]. By providing tactile sensations, such as resistance when lifting virtual objects, haptic devices enhance motor learning and skill acquisition. Patients can engage in immersive exercises that mimic real-world tasks, such as picking up objects or walking through a virtual environment, encouraging recovery through realistic and engaging activities.

Beyond motor rehabilitation, XR is also used in cognitive rehabilitation for individuals with traumatic brain injury (TBI) or neurodegenerative diseases such as Alzheimer's. Virtual environments provide patients with tasks that mimic everyday challenges, such as navigating a virtual store or solving a puzzle, while multimodal sensors monitor their cognitive performance. The system can adjust the difficulty level based on the patient's responses, creating a personalized rehabilitation plan that supports both cognitive recovery and patient engagement [48], [59].

### 3.1.4. Pain Management

XR is proving to be a powerful tool for pain management, particularly in addressing chronic pain. VR distraction therapy involves immersing patients in virtual environments that divert attention away from pain by engaging them in complex, immersive tasks. The use of multisensory feedback, such as visual, auditory, and tactile inputs, helps to reduce the perception of pain, with patients reporting significant pain relief during immersive VR sessions [60].

Additionally, XR-based virtual mirror therapy has shown promise in treating phantom limb pain in amputees. By creating the visual illusion of movement in the missing limb using full-body tracking, patients can "see" their limb moving in the virtual environment, which helps reduce pain [15], [61]. The combination of haptic feedback and EEG monitoring allows clinicians to better understand how pain perception can be modulated through virtual experiences, providing new avenues for non-pharmacological pain treatment.

### 3.2. *Educational Applications: Engaging Learning Through XR*

XR technologies are transforming the educational landscape by creating interactive and immersive environments that engage multiple sensory modalities, enhancing both learning and retention. By incorporating eye-tracking, hand-tracking, haptic feedback, and multimodal interfaces, XR brings abstract concepts to life and tailors the learning experience to individual needs. These innovations make learning more engaging and adaptable, promoting active participation and deeper comprehension in a wide range of subjects.

### 3.2.1. STEM Education

XR environments hold immense potential for STEM education, enabling students to visualize and manipulate complex scientific concepts that are often difficult to grasp through traditional learning methods. In subjects such as biology, physics, and chemistry, XR allows learners to interact with digital representations of biological structures, chemical reactions, or physical simulations. For example, students can explore the human body at a cellular level or perform virtual dissections, offering a more interactive approach to understanding anatomical structures [62].

Hand-tracking technologies facilitate this process by allowing students to manipulate digital models with natural gestures like pinching or grabbing, making learning more intuitive. Similarly, eye-tracking can monitor students' focus, providing feedback to instructors about which parts of the lesson are capturing attention and which are not, allowing real-time adjustments to optimize engagement [24].

In chemistry, for instance, students can perform virtual experiments by interacting with molecular models or balancing chemical equations in a dynamic environment, ensuring a hands-on learning experience without the risks associated with traditional lab settings. Haptic feedback further enriches the learning process by introducing tactile sensations, enabling students to "feel" chemical bonds breaking or forming, or to experience resistance when interacting with virtual objects, which enhances the understanding of abstract concepts.

3.2.2. History and Humanities

XR also plays a crucial role in enhancing learning in the humanities, particularly in history education. Through AR, historical artifacts and cultural landmarks can be superimposed onto physical spaces, allowing students to explore ancient civilizations or historical events from the classroom. For instance, AR applications can project virtual Roman ruins or medieval castles into the school environment, enabling students to walk through and interact with these reconstructions [63].

With full-body tracking, students can physically engage with historical events by participating in reenactments of key moments in history, such as battles or diplomatic negotiations. This immersive, hands-on experience not only enhances historical knowledge but also fosters empathy and critical thinking by placing students in the shoes of historical figures, encouraging them to understand the decisions and actions taken in different cultural and temporal contexts.

Beyond AR, VR simulations can recreate historical cities or cultural sites, such as ancient Athens or Renaissance Florence, offering students the chance to experience history in a way that traditional textbooks and lectures cannot provide. These multimodal experiences are particularly effective in fostering historical empathy and a deeper connection to past events, as students can explore these environments at their own pace, engaging their senses in a more comprehensive learning experience.

3.2.3. Multimodal Learning

Multimodal learning is one of the most powerful features of XR in education. By integrating EEG, eye-tracking, and haptic feedback, XR systems can create learning environments that dynamically adapt to the cognitive state of the student. For example, when students' brain activity is monitored through EEG, the system can adjust the lesson's pacing and complexity to match their cognitive load, enhancing both engagement and retention [64].

For instance, if a student's brainwave data indicates cognitive fatigue or frustration during a particularly challenging task, the XR system can reduce the difficulty or introduce more engaging elements to maintain focus. Conversely, if the system detects high levels of focus, it may introduce more complex tasks to sustain the student's engagement. Hand-tracking and haptic feedback further enhance this adaptive system by allowing students to physically interact with digital content, making abstract concepts tangible and easier to comprehend.

In a study by Parong and Mayer (2018) [65], students using XR for biology education showed significantly higher retention rates and engagement compared to those using traditional learning methods. The study emphasized how multimodal interactions, particularly through hand-tracking and eye-tracking, allowed students to manipulate virtual biological systems and understand complex processes more effectively.

3.2.4. Language Learning

XR is also making strides in language learning by providing immersive, contextualized environments where students can practice conversational skills in realistic virtual scenarios. By incorporating facial tracking and GSR, virtual tutors can adapt their tone and responses based on the emotional state of the learner, offering a more empathetic and personalized learning experience. For example, a learner struggling with anxiety during a virtual language conversation can be provided with calming cues or simpler dialogue, easing them into more challenging scenarios [66].

Through full-body tracking, students can engage in virtual role-playing scenarios where they interact with other avatars or digital agents, practicing both verbal and non-verbal communication. These virtual interactions foster deeper language comprehension and cultural awareness, as learners experience language use in real-world contexts, such as ordering food at a restaurant or navigating a new city [67].

3.2.5. Special Education

XR technologies also provide significant benefits for special education. By offering customizable and immersive learning environments, XR can cater to students with a wide range of learning disabilities, including dyslexia, ADHD, and autism. XR systems that integrate EEG and eye-tracking can monitor a student's cognitive load and focus, adapting the content in real-time to suit their individual needs. For example, students with ADHD may benefit from XR environments that adjust to their focus levels by minimizing distractions and providing more engaging, multimodal tasks [57], [68].

For students with autism spectrum disorder (ASD), XR environments can be designed to provide controlled social simulations where they can practice social interactions and communication skills in a low-pressure, customizable environment [21], [69]. These virtual experiences can be personalized using GSR and facial tracking to monitor the student's emotional state, ensuring that the scenarios remain supportive and adaptive to their comfort levels [70].

*3.3. Professional Training and Skill Development*

XR technologies are transforming professional training and skill development by offering highly immersive, realistic simulations that replicate real-world tasks in a safe, controlled environment. By leveraging multimodal technologies such as hand-tracking, haptic feedback, EEG, and GSR, XR environments allow for hands-on experiences, fostering skill development and retention. The ability to simulate complex scenarios, monitor physiological and cognitive states, and provide real-time feedback makes XR a powerful tool across industries, including healthcare, emergency response, manufacturing, and engineering.

3.3.1. Medical Training

In medical education, XR simulations enable students and trainees to practice complex procedures such as surgery with a high level of realism. Hand-tracking allows trainees to interact with virtual surgical tools, while haptic feedback simulates the tactile sensations of cutting tissue, suturing, or manipulating delicate organs. These immersive environments help develop motor skills that are crucial for real-world surgical procedures. Studies have shown that using XR for surgical training enhances learning and reduces the time it takes to acquire practical skills [71], [72].

EEG monitoring provides additional insights into the cognitive load of trainees during surgeries, helping instructors optimize training by adjusting the complexity of tasks based on mental fatigue and focus. This is particularly important in surgeries that demand long periods of sustained concentration. Haptic feedback and EEG integration ensure that the physical and cognitive aspects of learning are addressed, enhancing both skill acquisition and cognitive endurance [73].

3.3.2. Emergency Response

XR has also proven invaluable in training first responders for high-pressure situations such as natural disasters, fires, and mass casualty events. These environments simulate crisis scenarios that allow responders to practice search and rescue operations, medical triage, and decision-making in real time. Full-body tracking ensures that trainees' physical movements within the virtual environment are realistic, which is essential for learning proper techniques in confined or dangerous spaces [74].

To enhance this training, heart rate monitoring and GSR track the physiological stress responses of trainees, helping instructors understand how well they are managing high-stress situations. Studies show that physiological feedback can improve performance by teaching responders how to control stress in real-world scenarios (Andersen et al., 2018). These technologies make it possible to train for complex, dangerous situations in a way that is both safe and effective, helping to build confidence and competency.

3.3.3. Manufacturing and Engineering

In manufacturing and engineering, XR technologies facilitate virtual training in machinery operation, assembly processes, and maintenance, allowing workers to practice without the risk of damaging equipment or causing injury. Hand-tracking enables trainees to interact with complex machinery in virtual environments, simulating real-world gestures like turning knobs, pulling levers, or using specialized tools. Haptic feedback enhances this experience by providing tactile sensations, such as the resistance of materials or the texture of surfaces, offering a more comprehensive training experience [37], [75].

Eye-tracking is used to monitor where trainees focus their attention during assembly or repair tasks. This information helps instructors identify potential areas where further guidance is needed, improving the effectiveness of the training [75]. EEG sensors can monitor cognitive load, ensuring that tasks are presented in a way that challenges trainees without overwhelming them. Studies have shown that XR environments can reduce training time and improve retention in industrial settings by providing a more interactive and engaging experience [76].

3.3.4. Aviation Training

In aviation, XR simulations offer a realistic and safe environment for pilots to practice maneuvers, emergency responses, and routine operations. Hand-tracking technology allows pilots to interact with virtual cockpit controls, such as throttle levers, instruments, and flight sticks, while haptic feedback simulates the resistance and forces experienced during flight, such as turbulence or control adjustments. This technology provides pilots with the opportunity to practice handling complex situations without the risks associated with real-life flying [74].

EEG and heart rate monitoring are integrated into these systems to track cognitive and physiological responses during high-stress scenarios like engine failure or extreme weather conditions. This data allows instructors to assess a pilot's decision-making abilities under pressure and adjust training to improve stress management and focus [74]. XR environments also allow for repetitive practice, improving muscle memory and confidence in critical skills.

3.3.5. Corporate and Soft Skills Training

XR is also making a significant impact in corporate training, particularly for developing soft skills such as communication, leadership, and conflict resolution. In virtual environments, users can engage in realistic simulations of business meetings, negotiations, or customer service scenarios. Facial tracking and eye-tracking technologies allow avatars to mirror users' facial expressions and body language, making interactions more lifelike. This is particularly useful for leadership training, where participants practice managing teams or handling difficult conversations [39], [76], [77], [78].

For example, in customer service training, XR simulations place users in challenging client interactions, where hand-tracking and full-body tracking allow them to practice gestures and body language in response to virtual customers. EEG data is used to monitor emotional engagement and cognitive focus, ensuring that trainees remain engaged and responsive during interactions [53]. Studies have shown that this immersive approach improves interpersonal communication skills and emotional intelligence, which are critical in customer-facing roles [66].

3.3.6. Military and Defense

In military training, XR simulations are increasingly used to prepare soldiers for combat, strategic operations, and equipment handling. By integrating full-body tracking and haptic feedback, soldiers can engage in simulated battlefield environments where they practice maneuvers, weapon handling, and tactical coordination. The tactile sensations of haptic devices help replicate the experience of firing weapons or feeling explosions, enhancing the realism of the training [35], [37], [79].

Heart rate monitoring and GSR data track soldiers' stress levels during these simulations, helping instructors gauge their ability to manage fear and anxiety in high-pressure combat scenarios. This data provides valuable insights into how soldiers perform under stress, allowing for targeted feedback and the development of coping strategies [53], [66]. Research shows that XR military simulations significantly improve readiness and performance in real-world combat situations [73], [79].

*3.4. Arts and Entertainment: Immersive and Interactive Creativity*

In the realm of arts and entertainment, XR is enabling new forms of creative expression by integrating multimodal interactions that engage the user on multiple sensory levels. From interactive gaming to virtual galleries, XR offers immersive experiences that merge the physical and digital worlds.

Gaming: Multimodal technologies such as facial tracking, hand-tracking, and haptic feedback allow gamers to experience more realistic and emotionally engaging gameplay. Facial tracking replicates players' facial expressions on their avatars, enhancing social interaction and emotional immersion in multiplayer environments. Haptic feedback makes virtual actions more tangible, whether players are feeling the impact of an explosion or the sensation of handling objects training [35].

Virtual Art Galleries: XR allows artists to create and display virtual art installations that can be explored by users in fully immersive environments. Hand-tracking enables visitors to interact with artworks, while eye-tracking provides curators with insights into how audiences engage with different pieces [19], [30]. Haptic feedback can even simulate the texture of virtual sculptures or paintings, adding a tactile dimension to the art experience [31], [32], [34], [37].

*3.5. Public Health and Safety Training*

XR has the potential to revolutionize public health campaigns and safety training, providing an interactive and immersive way to educate the public and professionals about critical safety practices. By integrating multimodal technologies such as eye-tracking, hand-tracking, and haptic feedback, XR provides a comprehensive tool for training professionals and educating the public on critical health and safety issues. These technologies enable simulations that mimic real-world emergency and public health situations, allowing participants to experience scenarios in a controlled yet engaging manner, improving both retention and practical application of learned skills.

3.5.1. Disaster Preparedness

Disaster preparedness is an essential area where XR is proving highly effective. First responders, firefighters, and other emergency personnel can train in virtual disaster environments that simulate scenarios such as earthquakes, floods, or building collapses. These virtual environments are designed to replicate the chaos and unpredictability of real-life disasters, offering professionals the opportunity to practice making quick, informed decisions under pressure.

Full-body tracking plays a critical role in these scenarios by enabling realistic movement through virtual spaces. Trainees can navigate complex environments, such as collapsed buildings or hazardous areas, while haptic feedback provides tactile responses, allowing them to feel the weight of objects like fire hoses, rubble, or tools during search and rescue operations [31], [32], [34]. The integration of eye-tracking can further assess where trainees focus their attention during an emergency, offering insights into how they prioritize tasks, which can be used to optimize future training programs [19], [26].

XR also facilitates collaborative training by enabling multiple participants to interact in the same virtual environment. This capability is especially valuable for team-based emergency responses, where coordination and communication are critical. Real-time facial tracking can capture stress or confusion, offering instructors valuable data to refine their approach to team-based emergency management training[39], [76]. Through realistic and immersive training, XR can prepare professionals for a wide range of disaster scenarios, improving readiness and response capabilities [80].

### 3.5.2. Public Health Campaigns

XR can also be applied to public health campaigns, providing an interactive and engaging way to educate the general public about important health and safety protocols. These virtual environments allow individuals to experience critical health scenarios firsthand, such as responding to pandemics, natural disasters, or hazardous chemical spills. In such scenarios, participants can walk through the necessary steps for containment, protection, or emergency evacuation, all within a safe, simulated environment.

By using facial tracking and eye-tracking, XR platforms can monitor emotional engagement, ensuring that users remain focused and that the training resonates with them on an emotional level [19], [39], [63]. These features allow for personalized feedback and adaptation during the experience—if a user shows signs of stress or confusion, the system can alter the pacing or complexity of the simulation to ensure the message is effectively communicated.

For instance, during a virtual pandemic response simulation, eye-tracking data can reveal which areas of information (e.g., proper handwashing techniques or mask usage) capture the most attention, allowing public health authorities to focus campaigns more effectively [26]. Similarly, GSR or heart rate monitoring can track physiological responses to stressful scenarios, such as learning about quarantine measures or medical procedures during an outbreak, enabling authorities to design public health messaging that better addresses common anxieties or misconceptions [53], [66].

### 3.5.3. Workplace Safety and Compliance

XR environments also provide interactive workplace safety training for industries such as construction, manufacturing, and chemical processing, where hazards are prevalent, and proper safety measures are crucial. Through multimodal integration, such as hand-tracking and haptic feedback, trainees can practice handling machinery or dangerous materials safely [30], [34].

For example, a trainee in a virtual construction site may learn to operate heavy machinery using hand-tracking, while haptic feedback replicates the tactile sensations of interacting with physical controls [81]. These simulations help workers develop muscle memory and gain confidence before operating machinery in real life. Additionally, full-body tracking can simulate realistic physical movements in hazardous environments, such as navigating scaffolding or confined spaces, ensuring that workers are fully prepared to adhere to safety protocols [79], [82].

In chemical processing plants, XR can simulate hazardous material spills, guiding employees through containment and cleanup procedures without exposing them to actual danger. Eye-tracking data can be used to assess whether trainees are properly following protocols, while GSR data can measure stress responses to simulate real-world urgency and pressure [26], [53].

## 3.6. Retail and E-Commerce

XR technologies are reshaping the retail landscape by providing immersive, interactive shopping experiences that bridge the gap between physical and online stores. Through the integration of hand-tracking, eye-tracking, haptic feedback, and full-body tracking, XR environments allow users to explore virtual shopping spaces, interact with products in more intuitive ways, and receive personalized, data-driven experiences.

### 3.6.1. Virtual Shopping

Virtual shopping environments in XR allow customers to explore virtual stores where they can interact with products as if they were physically present. Instead of relying on static images, shoppers can manipulate 3D product models using hand-tracking to rotate items for a 360-degree view or zoom in to inspect details like texture or design. This enhanced interaction provides a more engaging shopping experience, which is especially valuable for products that customers typically prefer to see and touch before purchasing, such as electronics, furniture, or clothing [83].

In virtual clothing stores, full-body tracking enables customers to "try on" outfits using realistic avatars that mirror their movements. This allows customers to visualize how clothing items fit and move on their body, reducing uncertainty about size and appearance before making a purchase. Eye-tracking technology provides retailers with valuable insights into customer behavior, helping them understand which products or parts of the store layout attract the most attention, enabling data-driven optimization of store designs and product placements [84].

3.6.2. Product Customization

XR also enables AR-enhanced retail applications, allowing customers to customize products directly within their own physical environments. For example, AR apps for furniture retailers allow users to visualize virtual furniture in their real-world space, adjusting color, size, or style to see how it fits into their home decor. Haptic feedback further enriches this experience by providing tactile sensations that simulate the texture or weight of virtual items, making the interaction more tangible and realistic [85].

Similarly, in industries like fashion or automotive, users can interact with virtual models of products to customize details such as fabric, color, or features. These virtual models, combined with real-time customization tools, allow customers to create tailored products and preview them in their intended settings, increasing their confidence in purchasing decisions. This personalized approach is key in promoting customer satisfaction and reducing return rates, which is a common challenge in online retail [86].

3.6.3. Enhanced Customer Insights

Beyond improving the user experience, XR technologies like eye-tracking offer significant benefits for retailers by collecting detailed behavioral data. This data provides insights into how customers interact with the store environment and products, revealing patterns in attention, interest, and decision-making. Retailers can use this information to personalize the shopping experience further, adapting product recommendations, store layouts, and marketing strategies based on customer preferences [86].

For example, eye-tracking can reveal which products capture the most attention and how long a customer focuses on certain items. This information enables retailers to optimize product placements or highlight specific features that resonate with customers. The integration of biometric feedback such as GSR can also provide deeper insights into customers' emotional responses to products, allowing retailers to tailor their messaging and user interface to match the emotional engagement of shoppers [87].

*3.7. Architecture and Urban Planning*

XR technologies provide architects, urban planners, and construction professionals with the ability to visualize and interact with virtual models of buildings or urban landscapes before they are constructed. These technologies, integrating hand-tracking, haptic feedback, eye-tracking, and full-body tracking, provide architects, urban planners, and construction professionals with more efficient and dynamic ways to design, collaborate, and present their work.

3.7.1. Virtual Walkthroughs

One of the most transformative applications of XR in architecture is the ability to conduct virtual walkthroughs of buildings or urban projects using MR [88]. These virtual walkthroughs allow architects to place digital models of proposed designs directly into physical spaces, giving stakeholders the opportunity to experience the building before it is constructed. Clients can explore the layout, lighting, and spatial arrangement in a way that goes far beyond traditional blueprints or 3D renderings.

Through eye-tracking and facial tracking, architects can gather valuable feedback on client reactions, such as which areas of the design are drawing the most attention or eliciting emotional responses [88]. This data helps architects refine their designs to meet client expectations, creating spaces that are not only functional but also emotionally resonant. For example, if clients spend more time focusing on particular features like open spaces or natural lighting, architects can emphasize or adjust these aspects accordingly.

3.7.2. Collaborative Design

In addition to immersive visualization, XR enhances collaborative design processes, allowing multiple professionals to work together on the same project from different physical locations. Through hand-tracking and full-body tracking, architects and engineers can manipulate 3D models in real-time, using intuitive gestures to adjust designs, place virtual tools, or test structural components [89]. This ability to interact with digital models in a shared virtual space fosters more efficient collaboration and innovation, especially in large-scale urban planning projects that require input from various disciplines.

For example, urban planners can simulate how new buildings will interact with existing infrastructure, analyzing factors such as traffic flow, environmental impact, and public accessibility. With haptic feedback, users can feel the texture and resistance of building materials, providing a tactile layer to the design process [34]. This is particularly valuable in construction, where realistic feedback on materials can help professionals plan how different elements will come together in the real world.

3.7.3. Infrastructure and Sustainability Planning

XR also plays a significant role in infrastructure development and sustainability planning. Architects and urban planners can simulate energy efficiency, material use, and environmental impact before construction begins. By incorporating sensor data into these simulations, they can predict how buildings will perform in various conditions, such as extreme weather events, optimizing designs for sustainability and resilience [90].

Through AR overlays, urban planners can visualize proposed infrastructure changes, such as new roads, bridges, or public transit systems, in real-world environments, allowing for better community engagement and feedback [91]. For instance, residents can use AR applications to see how new developments will affect their neighborhood, giving them the opportunity to provide input before construction starts.

3.7.4. Safety and Risk Assessment

In architecture and construction, safety is a critical concern, and XR provides valuable tools for conducting safety and risk assessments. By creating virtual simulations of construction sites, professionals can identify potential hazards, test safety protocols, and train workers in a controlled environment [81]. For example, full-body tracking can simulate how workers will navigate a site, while haptic feedback allows them to feel the weight and resistance of safety gear or equipment [32], [35], [37]. These simulations help prevent accidents and ensure that safety measures are fully understood and implemented before workers enter a real-world site.

*3.8. Sports Training and Performance Analysis*

XR technologies are increasingly utilized in the sports industry, providing athletes and coaches with immersive tools for performance analysis and skill development. By integrating modalities such as full-body tracking, heart rate monitoring, GSR, and EEG, athletes can receive personalized training experiences. These technologies offer real-time feedback, helping to refine techniques, improve mental focus, and optimize both physical and cognitive performance [92], [93].

3.8.1. Skill Development

In XR sports training, athletes can engage in realistic simulations of their sport, allowing them to practice specific skills such as a tennis serve, golf swing, or soccer kick. Hand-tracking and haptic feedback are critical for providing realistic sensations during these exercises. For instance, when practicing a tennis serve in a virtual environment, haptic feedback can replicate the tension of the racket strings and the impact of the ball, thereby helping athletes to improve their precision and consistency [34]. Studies have shown that haptic feedback significantly enhances motor skill development by enabling athletes to "feel" the interaction between themselves and virtual objects, thus creating more naturalistic and effective training environments [30], [32].

Moreover, full-body tracking plays a vital role in capturing every aspect of the athlete's movement [92]. This allows coaches and trainers to analyze motion in great detail, identifying areas for improvement. For instance, motion capture data can be used to identify inefficiencies in an athlete's technique, such as improper body alignment during a sprint or incorrect posture during a weightlifting movement. By providing instant feedback on movement patterns, XR enables athletes to make real-time adjustments, which accelerates skill acquisition and minimizes the risk of injury [93].

3.8.2. Mental Training

Beyond physical skills, mental training is crucial in sports, particularly in high-pressure situations. XR tools that incorporate GSR, EEG, and eye-tracking technologies help monitor an athlete's cognitive load and stress levels during training. For instance, GSR sensors detect physiological arousal associated with stress, enabling coaches to understand how athletes respond to intense moments during training sessions or simulated competitions [46]. In sports such as archery, tennis, or golf, where focus and emotional regulation are critical, coaches can use this data to introduce mental conditioning techniques, such as mindfulness exercises or breathing techniques, to help athletes manage their stress more effectively [92].

Additionally, EEG technology is employed to monitor brain activity, which offers insights into an athlete's level of focus, fatigue, and mental engagement [93]. For instance, athletes training for sports like Formula 1 racing or esports—where reaction time and mental acuity are vital—can benefit from EEG-based feedback systems that track how well they concentrate under high-speed, high-pressure conditions. This data can be used to adjust training strategies, ensuring athletes are mentally sharp when it matters most.

Eye-tracking technology is also highly effective in analyzing athletes' visual attention and decision-making processes. In team sports, such as soccer or basketball, where spatial awareness and quick decision-making are essential, eye-tracking systems can analyze where players focus their attention during gameplay. For example, an eye-tracking study in soccer might reveal that an athlete is fixating too long on the ball rather than scanning the field for passing options, leading to improved training drills that emphasize peripheral awareness and faster decision-making [94].

### 3.8.3. Performance Analysis

XR systems not only help athletes improve their skills but also serve as powerful tools for performance analysis. By combining full-body tracking, heart rate monitoring, and biometric data, XR systems can provide a comprehensive understanding of how athletes perform in real-time [92]. For instance, heart rate variability (HRV) can be used to measure an athlete's physiological response to stress, giving insights into their cardiovascular health and recovery capacity. Coaches can use this data to optimize training regimens by ensuring athletes are exercising at the right intensity levels, adjusting workloads to prevent overtraining and maximize performance.

In endurance sports like cycling or running, XR systems can simulate race conditions, including terrain, wind resistance, and competition, allowing athletes to practice in highly controlled environments. EEG and GSR technologies track mental and emotional engagement during these simulations, ensuring that athletes are not only physically but also mentally prepared for competitions. This holistic approach to training ensures that athletes can refine their technical skills while also enhancing their mental toughness and emotional resilience [46].

### 3.9. *Remote Work and Virtual Offices*

As remote work becomes increasingly widespread, XR technologies are reshaping how businesses operate, offering immersive, interactive virtual office spaces. Through the integration of multimodal interfaces—including eye-tracking, hand-tracking, facial tracking, and full-body tracking—virtual offices now provide a high level of engagement, collaboration, and realism [95]. These innovations bridge the gap between physical office spaces and remote work, creating an environment where employees can collaborate as if they were in the same room, regardless of their physical location.

### 3.9.1. Virtual Meetings

In XR-enabled virtual offices, meetings feel more interactive and natural than traditional video conferencing. Facial tracking allows avatars to replicate the subtle nuances of facial expressions, capturing emotions and enhancing communication [94]. Hand-tracking enables users to manipulate digital objects, making brainstorming sessions more interactive by allowing participants to move virtual sticky notes or interact with 3D models [72], [85].

Eye-tracking offers another layer of engagement by allowing facilitators to monitor attention levels and adjust their presentations in real time. For example, in a virtual meeting, the system can track where participants are looking and provide feedback to the speaker on which visual elements are capturing attention, ensuring that presentations remain engaging [19]. Full-body tracking further adds to the immersive experience, allowing users to navigate virtual environments and interact with virtual objects more naturally [75], [95].

### 3.9.2. Remote Collaboration

Beyond meetings, XR offers extensive tools for remote collaboration, particularly in industries requiring detailed interaction with 3D models, such as engineering, architecture, and marketing. Hand-tracking allows team members to manipulate virtual prototypes or CAD models in real time, while haptic feedback provides tactile sensations that mimic the feel of physical materials [34], [72]. This technology allows remote collaborators to work on the same virtual project, make adjustments, and provide feedback as though they were physically present together.

Full-body tracking enhances collaborative experiences, allowing users to move freely in the virtual space, which is particularly useful for large-scale projects like building designs or landscape architecture [89]. Additionally, EEG and GSR sensors monitor participants' cognitive and emotional states, ensuring that workloads and tasks are adjusted to optimize performance [43]. For instance, in high-stress tasks, such as emergency response training, GSR data can prompt instructors to introduce relaxation exercises or modify the task to reduce stress [42].

### 3.9.3. Training and Onboarding

XR is revolutionizing training and onboarding processes in virtual offices by providing immersive, hands-on learning experiences. New employees can participate in virtual orientation sessions, explore digital replicas of office spaces, and interact with company resources using hand-tracking and eye-tracking technologies [85], [95]. Trainers can monitor cognitive engagement through EEG and adjust the training in real time, ensuring that trainees are neither under- nor over-stimulated [64]

Moreover, virtual office tours allow new hires to navigate the workspace and familiarize themselves with team members and company culture, making the transition smoother even when they are not physically present in the office [85]. This virtual onboarding reduces the time and resources traditionally spent on in-person orientations and helps employees feel more connected to the organization from the outset.

3.9.4. Future of Remote Work with XR

As XR technologies evolve, remote work and virtual offices will continue to offer immersive, high-performance environments that replicate and, in some cases, improve upon physical office spaces. Multimodal interfaces—such as haptic feedback, facial tracking, and EEG monitoring—will enhance remote collaboration, allowing for deeper engagement, more intuitive interaction, and improved team dynamics [2], [94]. AI-driven XR environments will further personalize the user experience by adjusting to users' cognitive and emotional states in real-time, ensuring optimal performance and engagement [2].

The seamless integration of physical and virtual workspaces will enable organizations to operate more effectively on a global scale, overcoming the barriers posed by geographical distance and time zones. XR will become a cornerstone of the future workplace, making remote work more accessible, inclusive, and productive. [95]

*3.10. Museums and Cultural Heritage Preservation*

XR technologies offer groundbreaking solutions for the preservation of cultural heritage and the creation of immersive museum experiences. By integrating multimodal interaction—such as eye-tracking, hand-tracking, and haptic feedback—museums can create more interactive, educational, and engaging experiences for visitors. Moreover, these technologies help preserve and showcase cultural artifacts and sites that may be inaccessible or endangered, allowing broader audiences to appreciate and learn about history and heritage [96], [97].

3.10.1. Virtual Museums

Virtual reality enables museums to go beyond traditional displays by creating immersive exhibits where visitors can explore digitized artifacts in ways previously unimaginable. For example, visitors to a virtual museum can use hand-tracking to "pick up" and manipulate 3D replicas of ancient relics, providing a more intimate and interactive experience. This method allows users to inspect objects from all angles, gaining a deeper understanding of the artifact's craftsmanship and details. Haptic feedback further enhances this experience by simulating the texture, weight, or material of these virtual objects, giving users the sensation of holding these artifacts in their hands, thus bridging the gap between physical and digital engagement [98], [99].

This multimodal interaction can significantly enrich educational experiences. For instance, a visitor in a virtual museum could use eye-tracking to interact with dynamic content that adapts based on their focus, offering personalized information or guided tours depending on where they direct their gaze. Such features cater to the user's learning pace and interests, promoting deeper engagement with the exhibit [96], [99].

3.10.2. Cultural Heritage Preservation

In the domain of cultural heritage preservation, XR technologies are being used to create detailed digital records of endangered or inaccessible cultural sites. Through 3D scanning and reconstruction, museums and cultural institutions can offer virtual tours of sites that are either too remote, fragile, or in danger of being lost to time. For example, endangered cultural sites, such as ancient temples or archaeological ruins, can be digitally preserved and reconstructed using XR, providing a detailed and immersive exploration of these landmarks for future generations [97], [100], [101]. Visitors can walk through digitally reconstructed environments using full-body tracking, exploring ancient cities, pyramids, or even World Heritage sites as they once appeared in their prime. Eye-tracking can be employed to gather data on which aspects of the environment capture the most attention, allowing curators to better understand how visitors engage with these reconstructions. This data can then be used to further refine and enhance the educational value of these digital exhibits [96], [101].

Moreover, XR can play a critical role in reconstructing destroyed or damaged cultural heritage. For example, following the destruction of sites due to war, natural disasters, or environmental factors, XR tools can assist in virtually rebuilding these locations, allowing historians, archaeologists, and the public to virtually explore and study them. In many cases, the digital reconstruction may be the only way to experience such sites in the future [97].

### 3.10.3. Multimodal Engagement in Museums

XR's multimodal interaction capabilities offer innovative ways for museums to engage with diverse audiences. With hand-tracking, visitors can interact more intuitively with virtual displays, while haptic feedback creates a sensory connection to the digital content, making it feel more tangible and real. Eye-tracking technology enhances educational engagement by identifying the most compelling parts of an exhibit, providing curators with valuable insights into visitor behavior, while simultaneously creating adaptive and personalized tour experiences [102].

The ability of XR to merge physical and digital experiences offers museums the potential to redefine how visitors engage with cultural heritage. The fusion of these multimodal technologies ensures that cultural knowledge is not only preserved but also made more accessible, interactive, and meaningful to future generations.

## 4. Potential Risks and Ethical Challenges of XR and the Metaverse

While the potential applications of XR technologies and the Metaverse offer groundbreaking advancements in education, healthcare, professional training, and entertainment, their rapid evolution also introduces a range of risks and ethical challenges (see Table 3). As users increasingly integrate their cognitive and physical selves with immersive digital environments, new concerns emerge regarding privacy, safety, and well-being. These technologies, while transformative, pose risks such as cybersickness, addiction, and cyber harassment, alongside deeper concerns about the collection and use of sensitive biometric, behavioral, and emotional data. As XR and the Metaverse continue to evolve, it becomes crucial to address these ethical challenges and ensure responsible development, balancing the immense opportunities with the potential dangers they bring.

**Table 3.** Potential Risks and Ethical Challenges in XR and the Metaverse

| Risk | Primary Concern | Implications | Examples |
|---|---|---|---|
| Cybersickness | Physical discomfort due to sensory conflict between visual input and lack of physical movement | Decreased engagement, reduced productivity, and hindered adoption of XR technologies in training or education | Users experience dizziness, nausea, and headaches during long VR training sessions, limiting effectiveness |
| Addiction & Dissociation | Psychological dependence and detachment due to immersive virtual environments | Neglect of real-world responsibilities, reality confusion, and potential mental health deterioration | Users spend excessive time in virtual spaces, leading to poor academic/work performance and reality disorientation |
| Cyber Harassment & Bullying | Anonymity and lack of consequences in virtual environments lead to harmful behaviors | Increased emotional distress, anxiety, and psychological harm due to virtual harassment or bullying | Virtual groping or abusive language in social XR spaces, with victims feeling violated and emotionally distressed |
| Data Privacy & Security | Collection of sensitive biometric, behavioral, and emotional data presents risks of misuse or hacking | Unauthorized access to personal data, identity theft, surveillance, and manipulation of user behavior | Biometric data like eye-tracking or EEG used to profile users and influence decision-making without their knowledge |
| Intense Advertising | Hyper-targeted and manipulative advertising based on user data | Manipulation of users' emotions and cognitive states, commercial exploitation, and over-commercialization of virtual spaces | Real-time emotional responses used to deliver targeted ads that exploit vulnerable moments, leading to compulsive purchases |
| Manipulation of Public Opinion | Misinformation and deepfake content spread through immersive experiences | Influence on political or social views, creating echo chambers and distorting users' perceptions of reality | Deepfake avatars of political figures delivering fake speeches in virtual environments, confusing |

| Risk | Primary Concern | Implications | Examples |
|---|---|---|---|
| | | | users about real-world events |
| Physical Health Concerns | Strain from prolonged use of XR headsets and repetitive movements | Eye strain, musculoskeletal problems, and sedentary lifestyle leading to long-term health issues | Users experience headaches, discomfort from extended headset use, or poor posture during VR interactions |
| Digital Divide & Inequality | Economic and infrastructure barriers limit access to XR technologies | Widening social and economic gaps, unequal access to education, healthcare, and workforce opportunities | Lower-income communities unable to access XR-enhanced education, resulting in disadvantaged students |
| Regulatory Challenges | Lack of comprehensive legal frameworks governing data privacy, content moderation, and intellectual property | Difficulty in enforcing user protections across decentralized and global XR platforms | Jurisdictional issues in handling virtual harassment or disputes over digital assets in cross-border XR environments |

*4.1. Cybersickness*

Cybersickness, a form of motion sickness, presents a significant challenge in XR, particularly in VR environments. It results from sensory conflict between the user's visual perception of movement and the lack of corresponding physical movement, causing symptoms like dizziness, nausea, headaches, and disorientation [103], [104]. These symptoms can limit user engagement with XR, particularly in educational and professional settings where prolonged immersion is required [105], [106], [107].

Beyond discomfort, cybersickness impairs cognitive performance and productivity in tasks requiring precise interactions or intense focus, such as virtual medical training [108]. If users associate XR with discomfort, they may reduce their use, particularly in settings that require long-term immersion [106], [107], [109]. Additionally, susceptibility varies widely; while some users experience mild symptoms, others may be unable to use VR technology without significant discomfort [23], [104], [110], [111].

4.1.1. Prevalence and Contributing Factors

Research indicates that between 20% to 80% of users experience cybersickness depending on the VR system, content, and individual sensitivity [103], [104], [106], [107], [110]. Contributing factors include frame rate, latency, environmental complexity, and fast-paced motion in virtual environments. Individual factors such as age, motion sickness susceptibility, and vestibular disorders also affect user experiences, complicating efforts to create universal solutions.

4.1.2. Technological and Design Solutions

Improving XR system performance by increasing frame rate and reducing latency helps reduce sensory conflict and cybersickness [107], [112]. Additionally, optimizing virtual environments—limiting rapid movements, simplifying camera angles, and incorporating smoother transitions—can minimize symptoms [103], [109]. "Comfort modes" that reduce the field of view during fast motion are also effective [107], [112]. Real-time physiological monitoring through sensors like eye-tracking and heart rate detection offers potential for managing cybersickness. If elevated heart rates indicate motion discomfort, the system can automatically adjust movement or visual stimuli to reduce symptoms [105], [107].

*4.2. Mental Health and Behavioral Risks: Addiction & Dissociation*

The immersive and engaging nature of XR technologies, especially the metaverse, presents significant mental health risks, such as addiction and psychological detachment. As these environments increasingly blur the lines between the virtual and physical worlds, users may experience challenges like compulsive engagement, disconnection from reality, and even psychological dissociation. These behavioral risks are heightened by the depth of involvement users can have with their avatars, social interactions, and achievements within virtual spaces [113], [114].

4.2.1. Addiction and Over-Immersion

XR's ability to provide endless exploration and social engagement creates a fertile ground for compulsive behaviors similar to those observed in internet and gaming addiction [115]. Users may become overly invested in their virtual lives, achievements, or social relationships, leading them to neglect real-world responsibilities such as work, education, or personal relationships [116]. The continuous availability of virtual environments, combined with reward systems, social validation, and in-game currencies, reinforces addictive behaviors [117].

The psychological toll of addiction manifests in various ways, such as poor academic or job performance, strained familial connections, and a diminished sense of real-world fulfillment [118]. Users seeking an escape from real-life stressors, such as personal or mental health issues, may immerse themselves in the metaverse as a coping mechanism, exacerbating conditions like anxiety or depression [119]. This pattern mirrors traditional forms of digital escapism, where users retreat into virtual worlds, only to find their mental health deteriorating over time [120], [121].

4.2.2. Dissociation and Psychological Detachment

The hyper-immersive nature of XR environments can also lead to more severe psychological outcomes, such as dissociation and reality confusion. Users may struggle to distinguish between their virtual and real-world experiences, particularly after long or repeated sessions in highly immersive environments [17]. This blurring of boundaries can lead to cognitive disorientation, memory lapses, and difficulty reintegrating into real-world activities [4].

A more extreme manifestation of dissociation is depersonalization, where users feel detached from their physical bodies or sense of self. In XR environments, where avatars can take on entirely different appearances and actions, users may start identifying more with their virtual personas than their physical identities. This can result in a diminished sense of control over their actions or a feeling of observing themselves from an external perspective [40], [113]. Such experiences can lead to emotional distress, particularly for those already prone to mental health issues, such as PTSD or anxiety disorders [122].

4.2.3. Mitigation Strategies for Mental Health Risks

To mitigate the risks of addiction and dissociation, XR developers and policymakers must implement design principles that promote healthy engagement. Time limits, break reminders, and "wellness" notifications can help users manage their time in virtual spaces and prevent over-immersion [121]. Educating users about responsible use and encouraging balance between virtual and real-world activities is crucial, particularly for vulnerable populations like adolescents and individuals with mental health conditions [118].

Additionally, XR systems should integrate monitoring features that can detect early signs of psychological distress, such as prolonged usage patterns or heightened emotional responses. These systems could provide real-time interventions, offering users the chance to take breaks or access mental health support when needed [4]. Grounding techniques, such as focusing on physical sensations or performing brief exercises, can also help users transition back into the real world after extended XR exposure [113].

Finally, clear guidelines on safe usage and transparency about the psychological risks of XR technologies must be established to empower users to make informed decisions. Providing resources on mental health support and creating policies that address the unique psychological challenges posed by XR environments will help ensure that these technologies are used responsibly [2].

*4.3. Cyber Harassment and Cyberbullying*

Cyber harassment and cyberbullying are growing concerns in XR and metaverse environments, particularly in virtual social spaces where users interact through avatars. The perceived anonymity and lack of real-world consequences in these immersive environments foster conditions conducive to harmful behavior. This reduced accountability, coupled with the decentralized nature of many virtual spaces, amplifies the risks of cyber harassment and bullying, making them pressing issues for the future of virtual interactions [123], [124]. As XR technologies become more widely adopted, addressing these issues is critical to ensuring the safety and inclusivity of these virtual environments.

4.3.1. Virtual Harassment and Emotional Distress

In XR environments, users are vulnerable to various forms of harassment, including virtual groping, unwanted interactions, and verbal abuse. The immersive nature of XR technology exacerbates these issues, as users often experience such violations as highly personal and immediate [125], [126]. This emotional intensity arises because avatars in virtual environments mirror users' physical movements and actions, making the harassment feel more real and invasive. For example, instances of sexual harassment in VR, where an individual's avatar is touched without consent, can provoke psychological trauma akin to real-world experiences of assault. The sense of presence in these environments amplifies the emotional impact of such actions, potentially leading to long-term emotional distress [117], [125].

4.3.2. Increased Aggression Due to Avatar Anonymity

The anonymity provided by avatars in XR allows individuals to engage in more aggressive behavior than they might in real life. This disinhibition effect leads to more abusive language, threats, and targeted harassment campaigns [124], [127]. In these virtual environments, users feel shielded from the repercussions of their actions, which can embolden bullies to act more viciously. The integration of multimodal feedback, such as haptic technology, further heightens the emotional and psychological toll on victims. For example, in VR spaces, the sensation of touch through haptic feedback can make unwanted interactions, such as virtual hitting or touching, feel more tangible and emotionally intrusive [117], [128]. This dynamic increases the psychological damage inflicted on victims, potentially leading to anxiety, depression, and other mental health issues [125].

### 4.3.3. Amplified Bullying Impact Through Multimodal Feedback

The integration of advanced technologies, such as facial tracking and haptic feedback, in XR environments can exacerbate the impact of cyberbullying. In traditional online spaces, harassment may be limited to text or voice interactions, but in XR, multimodal feedback creates a more immersive and realistic experience [129]. The ability to simulate physical sensations through haptics or detect facial expressions makes bullying interactions feel more personal, making it harder for victims to distance themselves emotionally from the experience. As a result, victims of bullying in these immersive environments are more likely to experience long-term psychological trauma, as the emotional weight of the harassment is intensified [127].

### 4.3.4. Challenges in Moderating XR Spaces

Moderating user behavior in XR environments presents unique challenges. Unlike traditional social media platforms, XR environments are vast, decentralized, and designed for real-time interactions [123]. This scale makes it difficult for moderators or algorithms to detect and respond to incidents of harassment or bullying promptly. Moreover, many XR platforms lack well-established governance structures, making it challenging to enforce rules or impose penalties for inappropriate behavior. The complexity of moderating these virtual spaces increases the likelihood that harmful behavior will go unnoticed or unpunished, leaving victims vulnerable to repeated harassment [124]. Furthermore, the lack of clear legal frameworks or recourse options for victims exacerbates the situation, as many users may not know where to turn for help when they experience abuse in XR environments [123].

### 4.3.5. Psychological and Social Consequences

The psychological impact of cyberbullying and harassment in XR environments mirrors the effects observed in traditional online bullying, but the immersive nature of XR can intensify these consequences. Victims of virtual harassment often report feelings of isolation, helplessness, and anxiety, which can lead to depression or suicidal ideation in severe cases [125], [127]. The emotional intensity of harassment in XR environments, combined with the public nature of many virtual interactions, can also lead to social exclusion. Victims may withdraw from virtual communities out of fear of further harassment, reducing their social engagement both online and in real life [124]. This withdrawal can have far-reaching effects on mental health, as social connections and community involvement are critical to emotional well-being [126].

### 4.3.6. Real-Time Monitoring and Reporting Systems

To effectively combat cyber harassment and bullying in XR environments, it is essential to implement real-time monitoring systems capable of detecting harmful behavior as it occurs. AI-driven moderation tools can scan conversations for abusive language and flag inappropriate physical interactions between avatars [123]. These systems should be complemented by robust reporting mechanisms that allow users to report incidents of harassment quickly and easily [117]. Such systems will enable moderators to respond to reports promptly and prevent further abuse. Additionally, integrating machine learning algorithms capable of identifying patterns of harassment could help platforms proactively intervene before situations escalate [129].

### 4.3.7. Establishing Clear Codes of Conduct

A foundational step in addressing harassment and bullying in XR is establishing clear, enforceable codes of conduct that outline acceptable behavior and the consequences for violations [124]. These guidelines should be made available to users upon entering virtual spaces and reinforced regularly to ensure everyone is aware of the expectations. Developers and platform operators should also provide clear instructions on how to report violations and what steps will be taken in response to reported incidents [117]. By promoting transparency and accountability, these codes of conduct can create a safer virtual environment.

### 4.3.8. Personal Boundaries and Safety Features

XR environments should incorporate personal boundary settings and safety features to give users control over their interactions. Personal boundary settings can prevent avatars from entering another user's virtual space, thereby reducing the risk of unwanted physical contact or harassment [123]. These features should also include easy-to-use tools that allow users to block, mute, or report offenders, enabling them to protect themselves in real-time. Providing users with the ability to customize their virtual safety settings ensures that they can maintain control over their virtual experiences and reduce their exposure to harmful behavior [117].

4.3.9. Collaboration Between Stakeholders

Addressing the issue of cyber harassment and bullying in XR requires collaboration between developers, policymakers, and legal authorities. Establishing governance frameworks that hold individuals accountable for their actions in virtual environments is crucial [124]. Legal recourse options should be available for victims, including the development of international standards for behavior in the metaverse. Governments, corporations, and advocacy groups must work together to create regulations that ensure XR environments are safe, inclusive, and free from harassment [123]. A collaborative approach will help create a framework for addressing the unique challenges posed by harassment in virtual spaces and ensure that those responsible for harmful behavior are held accountable.

*4.4. Data Privacy and Security Risks*

XR technologies collect extensive biometric, behavioral, and cognitive data to create immersive and personalized experiences, but this also introduces significant privacy and security risks. Misuse or improper handling of such sensitive data could result in exploitation for malicious purposes, including identity theft, unauthorized surveillance, and manipulation [130], [131].

4.4.1. Biometric and Behavioral Data Vulnerabilities

Biometric data, including eye-tracking, facial recognition, and EEG data, plays a critical role in enhancing immersion by adjusting virtual environments based on real-time user inputs. However, unauthorized access to this data could allow third parties to track a user's movements, facial expressions, or even brain activity, posing significant privacy concerns [19], [132]. The intrusion into personal autonomy could result in identity theft or unauthorized surveillance, exacerbating the ethical challenges XR technologies present [43].

Behavioral data, including interaction patterns and decision-making processes in virtual environments, allows the creation of detailed psychological profiles of users. These profiles can be exploited for targeted advertising, manipulation, or shared with third parties, often without the user's explicit consent [85], [130]. The immersive nature of XR can deepen this concern, making users more vulnerable to behavioral manipulation in contexts like political or commercial decision-making [133].

4.4.2. Cognitive and Affective Data Exploitation

XR systems can measure users' cognitive and emotional states using sensors like GSR and heart rate monitors, allowing real-time adaptation of virtual environments. This data can enhance user experiences but also raises ethical concerns about the potential for cognitive and emotional exploitation. For example, EEG data could reveal intimate information about a user's thoughts and emotional vulnerabilities, which could be used for manipulative advertising or other malicious purposes [42], [43], [130]. The commercialization of such sensitive data amplifies the ethical concerns, as users may not be fully aware of the depth of data being collected and its potential implications.

4.4.3. Hacking and Security Breaches

XR systems, due to their complexity and growing adoption, are increasingly attractive targets for cyberattacks. Data breaches could expose sensitive biometric, behavioral, and cognitive data, leading to unauthorized surveillance, identity theft, or manipulation of users' virtual environments [130]. In extreme cases, hackers could gain control of virtual environments to create hostile or traumatic experiences for users. The growing integration of brain-computer interfaces (BCIs) in XR adds an additional layer of risk, as unauthorized access to neural data could allow manipulation of users' thoughts or behaviors [64]. Moreover, the decentralized nature of many XR platforms makes it difficult to secure these systems, leaving users vulnerable to cyber threats [131].

4.4.4. Surveillance and Data Misuse by Corporations

Many XR platforms are operated by large corporations with extensive control over the data they collect. This raises concerns about the potential for corporate surveillance, where user data is collected and analyzed for commercial gain without explicit consent [133], [134]. Biometric and behavioral data can be used to predict user behavior, refine algorithms, and influence decision-making, raising ethical questions about the balance between user autonomy and corporate exploitation [131]. The lack of transparency about data collection practices further exacerbates these concerns [85].

4.4.5. Mitigating Privacy and Security Risks

To address these privacy and security risks, XR developers must adopt transparent data policies that clearly inform users about data collection, usage, and sharing practices. Robust consent mechanisms should be implemented, allowing users to opt in or out of specific data collection processes [131]. Enhanced encryption techniques and security protocols, such as multi-factor authentication, are essential to protect sensitive data from unauthorized access [130]. Regular security audits and vulnerability assessments should be conducted to ensure that XR systems are adequately protected against evolving cyber threats.

4.4.6. Ethical Guidelines and International Collaboration

The development of regulatory frameworks governing the ethical use of data collected by XR technologies is essential to ensure that only necessary data is collected and that user privacy is protected. Ethical guidelines should focus on data minimization and transparency, ensuring that data collection practices remain clear and accountable [131], [133]. Additionally, international collaboration is necessary to create standardized data privacy regulations across jurisdictions, ensuring that users' privacy rights are protected globally [134]. By fostering a unified global approach, governments and industry leaders can address the complex and interconnected privacy risks associated with XR technologies [130]

*4.5. Intense Advertising and Commercial Exploitation*

The immersive nature of XR technologies and the metaverse opens new avenues for advertisers to embed themselves deeply into the user's virtual experience. As users navigate digital spaces, they are no longer merely exposed to traditional forms of advertising but rather engage with hyper-targeted and personalized ads embedded seamlessly within their environments [134], [135]. This shift presents several significant ethical challenges, ranging from user manipulation to concerns over the commercialization of immersive spaces.

4.5.1. Hyper-Targeted and Manipulative Advertising

One of the most pressing concerns in XR advertising is the use of hyper-targeted ads. XR technologies collect vast amounts of personal data, such as eye-tracking, behavioral patterns, and even cognitive states measured by EEG and GSR sensors [26], [42], [63]. This data allows advertisers to tailor their messages with unprecedented precision, responding to users' real-time interactions, interests, and emotional responses. The ability to track where a user looks, how long they focus on a particular item, and even their emotional reactions allows advertisers to create highly personalized experiences [133], [135].

While personalization can improve the relevance of ads, it also opens the door to manipulative advertising practices. For example, a user's emotional state—whether they are excited, stressed, or tired—could be used to strategically deliver ads that exploit their vulnerabilities [64]. This is particularly concerning in scenarios where users may not fully realize the extent to which their behaviors and emotional responses are being monitored and used to influence their decisions. Research has shown that this type of advertising can blur the line between persuasion and manipulation, particularly when ads are delivered during heightened emotional states [133], [134].

In XR environments, the manipulation extends beyond static advertisements. Advertisers can create fully immersive virtual experiences that feel natural and non-invasive, allowing users to interact with products or services in ways that foster engagement and emotional attachment [85]. For instance, users might be encouraged to explore virtual stores where they can try on clothes through their avatars or interact with virtual items that mirror their real-world counterparts. While this could enhance the shopping experience, it also raises ethical questions about the manipulation of user preferences and behaviors [83].

4.5.2. Commercial Exploitation of Personal Data

A significant concern surrounding XR advertising is the commercial exploitation of personal data. XR platforms collect detailed biometric and behavioral data, which can provide deep insights into users' psychological profiles [26], [42], [136]. For example, a user's emotional responses to stimuli, attention patterns, and cognitive engagement levels can be tracked and used to inform advertising strategies [26]. This data could potentially be sold to third-party advertisers or used to create detailed consumer profiles, raising concerns about privacy and consent [137].

Moreover, the collection of affective data, such as GSR and heart rate monitoring, allows advertisers to tap into users' emotional states in real-time, influencing their behaviors at particularly vulnerable moments [42]. For example, when a user is feeling stressed or emotionally charged, an advertisement could be designed to provide comfort or relief, making it more likely that the user will respond positively to the message [43]. The risk here is that users may not fully understand the extent to which their emotional and cognitive states are being used to influence their purchasing decisions, leading to a form of subconscious manipulation [133].

The use of such detailed data also introduces the potential for misuse or exploitation, particularly when users are unaware of the full scope of data collection. In some cases, advertisers may push the boundaries of ethical behavior by using cognitive and emotional data to manipulate user behavior, fostering over-reliance on virtual environments or encouraging compulsive purchasing behaviors [134], [138].

4.5.3. Over-Commercialization and "Ad Fatigue"

As XR platforms continue to grow, there is a risk that these spaces will become over-commercialized [135]. The immersive nature of XR means that advertisements are no longer confined to traditional banners or pop-ups; instead, they can be embedded directly into the user's virtual environment, making them harder to ignore. This constant exposure to immersive ads could lead to "ad fatigue," where users become overwhelmed by the sheer volume and intensity of marketing messages [139].

In traditional online environments, users already face a high level of exposure to advertisements, which can lead to disengagement and frustration. This problem is likely to be magnified in XR, where users are fully immersed in their surroundings and unable to easily escape intrusive ads [133]. The risk is that the over-commercialization of XR spaces could diminish the quality of the user experience, leading to psychological stress, distraction, and a reduction in the overall enjoyment of the immersive environment [63].

Furthermore, as advertisers seek to monetize the metaverse and XR platforms, users may find themselves increasingly bombarded by immersive advertisements that demand their attention. This could lead to a deterioration of trust between users and XR platforms, particularly if users feel that their immersive experiences are being compromised by commercial interests [131]. The challenge for XR developers and advertisers alike will be to strike a balance between monetization and user satisfaction, ensuring that ads are integrated seamlessly and ethically into the virtual environment.

4.5.4. Mitigation Strategies for Ethical Advertising

To mitigate the risks associated with intense advertising and commercial exploitation in XR, several strategies can be employed. One approach is to ensure that advertising is transparent and non-intrusive. Users should have clear control over the types of ads they encounter, as well as the ability to opt out of data collection practices that could be used for hyper-targeted marketing [137]. Another important strategy is the implementation of ethical design principles in the creation of XR advertisements. Advertisers should avoid exploiting users' emotional states or cognitive vulnerabilities for commercial gain [42], [134]. Instead, XR platforms should focus on creating positive, relevant experiences that respect the user's autonomy and well-being [2]. Finally, regulatory frameworks and data protection laws must evolve to address the unique challenges posed by XR technologies. Governments and policymakers should work with XR developers to ensure that users' privacy is protected and that the exploitation of biometric, behavioral, and cognitive data is strictly regulated [131], [137]. By fostering transparency, accountability, and user control, XR platforms can create a safer and more ethical advertising ecosystem.

*4.6. Manipulation of Public Opinion and Information*

The immersive and persuasive nature of XR platforms presents new risks for the manipulation of public opinion and the spread of disinformation. As XR technology integrates into everyday life, virtual environments may become increasingly influential in shaping individuals' beliefs and behaviors. These platforms provide fertile ground for disinformation campaigns, where misleading or fabricated content can be seamlessly woven into immersive experiences, making it harder for users to differentiate between fact and fiction [134], [140]. The convergence of AI, virtual worlds, and real-time user interaction heightens the risk of psychological and social manipulation, potentially transforming XR spaces into tools for influencing public opinion on a large scale [141], [142].

4.6.1. False Realities and Misinformation

One of the most significant dangers posed by XR is the ability to create entirely fabricated environments that users may accept as real. Misinformation can be embedded within these virtual worlds, where users interact with distorted or false narratives, potentially reinforcing existing biases or creating new ones [134], [140]. Virtual experiences are highly immersive, and the emotional and cognitive impact they can have on users often exceeds that of traditional media. In such contexts, individuals may be exposed to "echo chambers," where their beliefs and opinions are reinforced by interacting with only those ideas that align with their pre-existing views, exacerbating political polarization [128], [142].

Misinformation in XR environments is more than just textual or visual; it is experiential, making it harder for users to critically evaluate the information presented to them [143].

4.6.2. Influence on Behavior

The immersive nature of XR has a profound psychological effect, allowing users to engage with information on a much deeper level than traditional media [4], [144]. In these environments, users may be more vulnerable to manipulation, as immersive experiences often elicit stronger emotional responses, which can significantly influence decision-making processes [113]. This makes XR an ideal platform for orchestrating campaigns designed to sway public opinion or influence individual behaviors [125]. Virtual reality simulations that portray emotionally charged or controversial content could manipulate users' emotions to create fear, loyalty, or anger, effectively shaping how they view real-world events or political issues [142]. The use of cognitive biases, such as emotional appeal and repetition, in immersive environments makes it easier for disinformation campaigns to alter perceptions and beliefs subtly.

4.6.3. Deepfake Avatars and AI-Generated Content

Another major concern is the use of deepfake technology and AI-generated content within XR spaces. Deepfakes, which use artificial intelligence to create highly realistic but fabricated videos or avatars, can be employed to impersonate public figures or trusted individuals [134], [142]. In an XR environment, malicious actors could create convincing deepfake avatars of politicians, celebrities, or community leaders to mislead users or manipulate public discourse. For instance, a deepfake avatar of a political figure could deliver a fabricated speech in a virtual rally, promoting false claims or disinformation [140]. The seamless integration of these fake avatars into XR environments makes it difficult for users to detect the deception. Furthermore, AI-generated content—such as speeches, articles, or social media posts—could be used to flood virtual spaces with misleading information, influencing how users perceive events, public figures, and political issues [134], [143](Westerlund, 2019).

4.6.4. Improving Media Literacy and Critical Thinking Skills

One of the most effective ways to combat the manipulation of public opinion in XR environments is by improving media literacy and encouraging critical thinking [145]. Users must be taught how to evaluate the credibility of virtual experiences and be given tools to distinguish between factual information and manipulated content. Educational programs focused on identifying disinformation in XR spaces can help users become more skeptical of immersive content, making them less susceptible to manipulation [141], [145]. Additionally, implementing fact-checking mechanisms within XR environments can provide users with real-time feedback on the validity of the information they encounter.

4.6.5. Authentication of Avatars and Content

Developers of XR platforms should prioritize creating systems for verifying the authenticity of avatars and digital content. Authentication tools that use blockchain technology or secure identification processes can help ensure that avatars representing public figures or institutions are legitimate, reducing the risk of deepfake impersonations [146]. These systems could include digital signatures that verify the origin of content and guarantee that public discourse is not being manipulated by AI-generated or fabricated personas. Additionally, moderation tools that detect deepfakes or manipulated content in real-time should be integrated into XR environments to mitigate the spread of disinformation [140].

4.6.6. Regulation and Collaboration

Governments, tech companies, and policymakers need to collaborate to develop regulations that address the risks of disinformation in XR environments. These regulations should focus on preventing the exploitation of immersive technologies for political or financial gain, holding companies accountable for the content shared on their platforms [141], [146]. Additionally, platforms should be required to have robust reporting mechanisms for users to flag misleading or harmful content. As XR platforms grow in popularity and influence, clear governance frameworks will be essential to ensure that public opinion is not manipulated through immersive and deceptive virtual experiences [134].

*4.7. Physical Health Concerns*

While XR technologies provide immersive experiences, they also pose physical health risks, particularly from prolonged use of headsets and controllers. These risks include eye strain, musculoskeletal discomfort, and sedentary behavior, all of which can lead to long-term health issues if not properly managed [147], [148].

4.7.1. Eye Strain and Musculoskeletal Problems

Extended XR use can lead to digital eye strain (or computer vision syndrome) due to the proximity of screens and the visual adjustments required for 3D content, resulting in headaches, blurred vision, and dry eyes [147], [148]. Simultaneously, repetitive hand movements, poor posture, and heavy headsets can cause musculoskeletal discomfort, such as neck, wrist, and shoulder strain [149], [150]. These risks are exacerbated in professional or gaming contexts where users perform repeated physical tasks for extended periods.

4.7.2. Sedentary Lifestyle and Physical Inactivity

Although some XR applications promote physical activity, many experiences, especially in entertainment and gaming, contribute to a sedentary lifestyle. Users often sit for long periods with limited movement, increasing the risk of obesity, cardiovascular disease, and metabolic disorders [149], [150]. The immersive nature of XR can lead users to lose track of time, further exacerbating physical inactivity.

4.7.3. Ergonomic Design and Break Management

To reduce physical strain, XR systems should prioritize ergonomic design with lightweight headsets, adjustable straps, and controllers that promote natural movement. Additionally, XR applications should include reminders for screen breaks to mitigate eye strain and encourage movement [147], [148]. Regular breaks and optimized display settings, such as adjusting brightness and contrast, can help prevent digital eye strain and musculoskeletal issues.

4.7.4. Promoting Physical Activity in XR

To counteract the sedentary nature of many XR applications, developers can integrate movement-based activities and physical challenges into virtual environments, even those designed for entertainment or social interaction [151]. Fitness applications that require users to move, stretch, or exercise while engaging with the virtual environment are an example of how XR can promote physical activity [150]. Additionally, adopting built-in features that monitor inactivity and prompt users to move or take physical breaks could help mitigate the health risks associated with prolonged sitting [149].

*4.8. Digital Divide and Inequality*

As XR technologies become more integrated into various sectors such as education, healthcare, and industry, concerns are growing that these advancements could exacerbate existing social and economic inequalities [152], [153]. The high cost of XR hardware and the necessary infrastructure may create significant barriers for lower-income populations, leading to a widening digital divide [95]. This divide could result in a two-tiered society where only those with the financial means can fully participate in and benefit from the metaverse and XR applications, further entrenching socio-economic disparities [90], [146].

4.8.1. Economic Barriers

One of the primary challenges is the substantial cost associated with acquiring XR hardware. High-quality VR headsets, AR glasses, haptic suits, and other peripheral devices often come with a hefty price tag that is unaffordable for many individuals and institutions, especially in low-income communities and developing countries [95], [146]. Additionally, the infrastructure required to support these technologies—such as high-speed internet, powerful computers, and ample physical space—adds to the financial burden [152]. This creates a significant barrier to entry, limiting access to those who can afford these expenses.

The economic barriers to accessing XR technologies risk creating a two-tiered system of digital experience [153]. Individuals and communities with financial resources can access enhanced educational tools, healthcare services, and professional opportunities through XR, while those without such means are left behind [152]. This disparity not only limits personal and professional development for those in lower-income brackets but also reinforces existing social and economic inequalities [95]. The unequal distribution of technological resources can lead to a cycle where disadvantaged groups become increasingly marginalized in a digitally-driven society.

4.8.2. Access to Education and Healthcare

XR has the potential to revolutionize education by providing immersive and interactive learning experiences. However, without equitable access, students from lower-income families or underfunded schools may miss out on these advancements, widening the educational gap [11]. Schools in affluent areas may integrate XR into their curricula, offering students cutting-edge tools for learning complex subjects through virtual laboratories, historical simulations, and interactive problem-solving environments [96]. Conversely, schools lacking resources may continue with traditional methods, potentially disadvantaging their students in terms of engagement and skill acquisition [152].

In healthcare, XR technologies are being utilized for patient treatment, medical training, mental health therapies, and rehabilitation programs [154]. These innovations can improve patient outcomes and make healthcare services more efficient [15]. However, the high costs associated with XR technologies may prevent underfunded hospitals, clinics, and healthcare providers in low-income or rural areas from adopting them. This could result in unequal access to high-quality healthcare, where only patients in well-funded systems benefit from the latest XR-enhanced medical treatments and services [64], [154].

4.8.3.  Impact on Workforce Development

As industries adopt XR technologies for training and operations, workers without access to these tools may find themselves at a disadvantage in the job market [155]. Companies may prefer or require employees who are proficient in XR technologies, creating employment barriers for those from lower socio-economic backgrounds who have not had the opportunity to develop these skills [90]. This can exacerbate unemployment and underemployment in already vulnerable communities, further entrenching economic disparities [152].

4.8.4.  Reducing Cost Barriers

To bridge the digital divide, efforts must be made to reduce the cost of XR hardware and make it more accessible [95]. Manufacturers and developers can work towards creating affordable XR devices without significantly compromising quality [11]. Subsidies, grants, and financial assistance programs provided by governments, non-profits, and private organizations can help lower-income individuals and institutions acquire necessary equipment [153]. Bulk purchasing agreements and partnerships with educational and healthcare institutions can also reduce costs [96].

4.8.5.  Improving Infrastructure Access

Expanding high-speed internet access and improving technological infrastructure in underserved areas is critical [152]. Investments in broadband expansion projects, particularly in rural and low-income urban areas, can provide the necessary connectivity for XR applications [90]. Public-private partnerships can be instrumental in funding and implementing these infrastructure projects, ensuring that more communities can participate in the digital economy [95].

4.8.6.  Public Access and Community Programs

Creating public spaces where XR technologies are available can help mitigate access issues. Libraries, community centers, and schools can serve as hubs where individuals can experience and learn about XR technologies without personal ownership [146]. Educational programs and workshops can introduce XR to broader audiences, fostering digital literacy and skills development across socio-economic groups [153].

4.8.7.  Policy and Regulatory Interventions

Government policies can play a significant role in addressing digital inequality [152]. Implementing regulations that promote fair pricing, prevent monopolistic practices, and encourage competition among XR providers can help lower costs [95]. Additionally, policies that support funding for technology in education and healthcare can ensure that institutions serving low-income populations are not left behind. International cooperation may also be necessary to address global disparities in XR access and to develop standards that promote inclusivity [153].

4.8.8.  Encouraging Inclusive Design

Developers and content creators should consider the diverse needs of users from different socio-economic backgrounds [96]. Designing XR applications that can operate on lower-spec hardware or that require less bandwidth can make these technologies more accessible [11]. Additionally, providing multilingual support and culturally relevant content can enhance the inclusivity of XR experiences [90], [152].

*4.9.  Regulatory Challenges and Governance*

The decentralized and rapidly evolving nature of XR technologies and the metaverse presents complex regulatory challenges. These challenges include data privacy, user safety, intellectual property, and content moderation, all of which require robust legal frameworks [146], [156]. As XR technologies integrate further into everyday life, the absence of clear regulations creates significant gaps in user protection, leaving critical areas unregulated [130].

4.9.1.  Lack of Regulation

XR environments collect vast amounts of biometric, behavioral, and cognitive data, but existing regulations, such as GDPR and CCPA, do not fully address the unique privacy and security risks posed by XR [131], [137]. The rapid pace of technological advancements has outpaced regulatory efforts, leaving sensitive data vulnerable to misuse, identity theft, or surveillance [146]. As XR becomes more widespread, regulators must act swiftly to establish comprehensive policies that protect user data while fostering innovation.

4.9.2.  Jurisdictional Issues

The global nature of the metaverse raises significant jurisdictional challenges. Users from different countries interact in shared virtual spaces, complicating legal accountability when issues such as harassment or fraud occur [123]. Determining which legal jurisdiction applies to virtual environments—whether based on the user's location, the platform's host country, or the perpetrator's location—creates legal ambiguity. Effective regulation will require international collaboration and the establishment of cross-border agreements [134].

4.9.3. Content Moderation

Moderating user behavior and content in XR presents difficulties due to the vast scale and decentralized nature of virtual spaces. Harmful content such as harassment or hate speech can be more damaging in immersive environments, but current governance models, including decentralized autonomous organizations (DAOs), struggle to effectively moderate these interactions in real-time [126]. Content moderation systems in XR must adapt to the immersive, interactive nature of the technology while balancing transparency and user participation [130].

4.9.4. Intellectual Property Concerns

The creation of virtual assets in XR spaces raises significant intellectual property (IP) concerns. Digital goods like avatars, virtual real estate, and artwork can be copied or resold without the original creator's consent, leading to disputes over ownership [141]. The rise of NFTs has provided a mechanism for proving ownership but does not always prevent unauthorized duplication. Adapting IP laws to address virtual asset protection in decentralized environments is critical to safeguarding creators' rights [146].

4.9.5. International Legal Frameworks

To address regulatory gaps, governments must collaborate on developing global legal standards for XR platforms. This could involve treaties that establish principles for data privacy, user protection, and IP rights across virtual environments, ensuring a unified approach to cross-border governance [134]. Such frameworks would enable better regulation of XR spaces and ensure consistency in legal accountability [146].

4.9.6. AI-Driven Content Moderation

Advanced AI systems offer a solution for content moderation in XR. Machine learning algorithms can detect and flag inappropriate behavior or harmful interactions in real-time, helping to create safer virtual environments [126]. AI can also identify patterns of misconduct before they escalate, ensuring more effective governance and user safety [130].

4.9.7. Strengthening IP Laws for the Metaverse

As the metaverse grows, governments must strengthen intellectual property laws to protect virtual assets. Clear guidelines on the creation, ownership, and distribution of digital goods, along with stronger penalties for infringement, will ensure better protection for creators. Regulatory bodies must also adapt to the rise of NFTs, ensuring that they provide adequate protection while preventing unauthorized replication of virtual assets [156].

**5. General Discussion**

This review examined the applications, risks, and ethical challenges associated with XR technologies, including VR, AR, and MR, across various domains. It is evident that cognitive processes now extend into digital environments, where tools, objects, and virtual spaces reshape how individuals think, learn, and solve problems. XR offers unprecedented opportunities for cognitive augmentation, particularly in fields such as education, healthcare, and professional training. However, alongside these advancements, the review highlights the complex ethical, psychological, and societal challenges that emerge, such as data privacy concerns, the risk of addiction, and exacerbated social inequalities. These challenges necessitate careful scrutiny and regulation to ensure that XR technologies are developed and deployed responsibly, safeguarding user well-being and ensuring equitable access across different populations.

*5.1. Applications of XR in Real-World Contexts*

The transformative potential of XR technologies is not only theoretical but is already being realized in practical applications across various fields. These technologies demonstrate significant promise in areas such as education, healthcare, and professional training by expanding cognitive capacities and enhancing user experiences through immersive environments.

5.1.1. Education and Training

One of the most prominent applications of XR technologies is in education, where they have revolutionized learning by offering immersive, interactive experiences that extend the traditional boundaries of instruction. In fields like science, technology, engineering, and mathematics (STEM), XR facilitates hands-on learning in virtual laboratories, where students can engage with complex concepts through virtual dissections, chemical simulations, and architectural designs [81]. These environments allow learners to offload cognitive tasks such as spatial reasoning and problem-solving to the virtual interface, enhancing their capacity to understand and retain information [67]. Medical training has also been significantly enhanced through XR applications. VR enables medical students to practice surgeries in a risk-free, controlled environment, receiving real-time feedback through haptic feedback systems [71], [82] This level of interaction, coupled with immersive simulations, has led to improved knowledge retention and skills acquisition, demonstrating the effectiveness of XR in training contexts [72].

5.1.2. Healthcare

Beyond education, XR is increasingly being used in healthcare, particularly for therapeutic interventions and medical training. XR-based cognitive-behavioral therapies (CBT) for anxiety, depression, and phobias allow patients to engage in exposure therapy within controlled virtual environments. These environments are personalized in real-time using biometric data, such as EEG and GSR readings, to monitor the user's emotional and cognitive states [4]. Such interventions have been shown to be particularly effective in treating specific phobias and PTSD [157]. Additionally, XR is being employed in pain management, where VR distraction therapy helps reduce chronic pain by immersing patients in calming virtual worlds [15]. The ability to manipulate patients' focus away from pain through immersive visual and auditory experiences provides a promising alternative to traditional pain management techniques. Moreover, XR technologies have significant potential in neuropsychological assessment, offering more ecologically valid testing of cognitive and motor functions. Tools like the Virtual Reality Everyday Assessment Lab (VR-EAL) enable assessments of prospective memory, executive function, and attention in realistic, daily-life environments, improving the accuracy of clinical diagnoses and rehabilitation compared to traditional methods [20], [22], [158]. The increased ecological validity and engagement of XR assessments make them valuable for both research and clinical practice.

5.1.3. Professional and Industrial Training

XR technologies have also proven to be invaluable in professional and industrial training, particularly in fields that require complex skills and decision-making under pressure. For example, in manufacturing and construction, workers can train in operating machinery and performing intricate tasks in a virtual space, reducing risk while enhancing proficiency [81]. Virtual environments simulate real-world conditions, allowing professionals to practice without the constraints of physical resources or the risks associated with errors. In emergency response training, XR is used to simulate high-stress scenarios, such as natural disasters or large-scale emergencies. First responders can rehearse their roles in virtual environments, enhancing their preparedness and decision-making capabilities. The ability to replicate these environments with real-time feedback using biometric monitoring adds a new dimension to training efficacy [80].

5.1.4. Entertainment and Cultural Heritage

In entertainment, XR has expanded the boundaries of user engagement. Video games and social platforms within the metaverse leverage XR to create fully immersive environments that respond to users' movements and emotions in real time [125]. These applications utilize advanced tracking systems such as eye-tracking and hand-tracking to mirror users' actions within the virtual world, creating a heightened sense of presence and interactivity [113]. Cultural heritage preservation has also benefited from XR technologies. Virtual museums allow users to explore historical artifacts and sites through immersive 3D reconstructions, with haptic feedback providing tactile sensations to deepen user engagement [98]. These experiences enhance accessibility to cultural education and provide new ways for users to interact with heritage, fostering a deeper connection to historical content [101].

*5.2. Ethical and Psychological Implications of XR*

While XR technologies offer significant cognitive and practical advantages, they also introduce several ethical, psychological, and social challenges. These issues require urgent attention, especially as XR technologies become more integrated into daily life and professional environments. The growing ubiquity of XR platforms raises questions about data privacy, psychological safety, social inequality, and the potential for long-term dependency on digital environments.

5.2.1. Privacy, Surveillance, and Data Ethics

The integration of biometric and behavioral data collection within XR environments raises significant privacy concerns. XR technologies often rely on tracking user movements, gaze, facial expressions, and even neural activity, all of which are highly sensitive data points [26], [148]. These data can be used to personalize experiences, optimize system performance, or monitor cognitive engagement, but they can also be exploited for commercial or political purposes. For example, eye-tracking data can reveal what users focus on, allowing marketers to target them with precision-based advertising, while brainwave monitoring could be used to manipulate user behavior by adjusting virtual environments to induce certain emotional or cognitive states [24], [133].

The integration of biometric and behavioral data collection in XR environments poses significant ethical and privacy challenges. Current frameworks, such as the GDPR, only partially address these issues, as they were not designed with immersive technologies in mind [131]. Eye-tracking, gait analysis, and even neural data, collected BCIs, can provide unprecedented levels of insight into users' cognitive and emotional states, which could be exploited for commercial or political manipulation [26], [159]. Moreover, AI-driven algorithms embedded within XR platforms can analyze these data in real time to adjust virtual environments, further heightening concerns about algorithmic manipulation of users' emotions and decisions [160].

As XR becomes increasingly integrated into the Metaverse—a persistent, shared digital universe—these privacy challenges are amplified. The Metaverse's reliance on real-time interaction and continuous data flow makes it more vulnerable to exploitation, particularly in environments where AI-driven algorithms process vast amounts of biometric data to optimize user experiences. This has raised concerns that the Metaverse could become an environment where users are subject to highly targeted manipulation, with few safeguards in place to ensure ethical data use [161]. New trends in the Metaverse, including decentralized governance and open standards, are promising but still insufficient to mitigate the risks posed by large-scale data collection and surveillance.

Ethical concerns surrounding informed consent in immersive environments must also be addressed. Users need to fully understand how their biometric data is being collected, processed, and potentially used. Without clear regulations, XR platforms risk becoming environments of mass surveillance, where users' privacy is compromised. Governments and regulatory bodies must work collaboratively to adapt existing laws like GDPR and develop new legal frameworks that account for the nuances of XR environments [130], [134]. These frameworks should emphasize user autonomy, ensuring that consent is informed and cognitive privacy is respected [133].

Additionally, there is a growing need to consider the implementation of these privacy and data protection standards in rapidly developing regions. While many countries in Europe and North America have comprehensive data protection laws, such as the GDPR, many developing regions lag behind in establishing such regulations [152]. As XR technologies become more widespread in these regions—particularly for educational, healthcare, and economic development purposes—the risk of data exploitation becomes greater in the absence of robust regulatory frameworks [162]. International cooperation is crucial to extend privacy protections to these regions and ensure that XR's deployment in developing areas follows ethical guidelines, safeguarding user privacy and autonomy.

To mitigate these challenges, policies must also promote digital literacy and ensure that users in developing regions understand the implications of data sharing in XR environments. This is especially important as these regions increasingly adopt XR for economic development and education. If left unregulated, the integration of XR into these areas could result in users being disproportionately vulnerable to data exploitation due to weaker legal protections and limited awareness of data privacy concerns [163].

5.2.2. Psychological and Behavioral Impacts

Prolonged exposure to XR environments may have significant psychological and behavioral effects, particularly as users become more immersed in these spaces. One major concern is the potential for XR addiction, where users spend excessive time in virtual worlds at the expense of their real-world responsibilities and relationships [125]. This is especially concerning in the context of entertainment-based XR, such as gaming or social platforms within the metaverse. The deep sense of presence and the compelling nature of these environments can make it difficult for users to disconnect, leading to negative impacts on mental health, such as anxiety, depression, or social isolation [121], [125]. Additionally, XR environments can lead to psychological dissociation, where users have difficulty distinguishing between the real world and virtual spaces [51], [125]. This blurring of reality can be particularly harmful for vulnerable populations, such as children or individuals with mental health conditions. Research has shown that immersive experiences can lead to heightened anxiety, confusion, or even hallucinations in some cases, especially if users are unable to reset their cognitive boundaries after prolonged use of XR systems [125], [164].

Another significant concern is the potential for harassment or inappropriate behavior within virtual environments. The sense of embodiment in XR can make interactions feel very real, and violations of personal space or harassment in virtual worlds can result in psychological trauma similar to that experienced in physical spaces [76], [126]. This underscores the need for robust systems of moderation and governance within XR platforms, ensuring that users can feel safe and protected in these environments.

*5.3. Societal and Technological Impacts: Bridging Realities and Digital Inequality*

5.3.1. Digital Divide and Access

As XR technologies advance, they may exacerbate existing inequalities related to digital access and literacy. While these technologies offer significant cognitive and professional advantages to those who can access them, they may also widen the gap between those with the resources and skills to engage with XR and those without [152]. Access to high-end XR systems often requires substantial financial investment, as well as access to fast internet, specialized hardware, and technological literacy [11]. These barriers could exclude low-income communities, underfunded schools, and developing regions from the cognitive benefits offered by XR [165].

Developing regions, in particular, may face additional challenges as they often lack the necessary infrastructure, such as broadband internet and affordable hardware, to fully leverage XR technologies. This digital divide is compounded by the uneven distribution of educational and technical resources. Consequently, the introduction of XR technologies without addressing these fundamental access issues could exacerbate existing social and economic inequalities, limiting opportunities for marginalized communities to benefit from innovations in education, healthcare, and training [162].

Moreover, as XR becomes more integrated into professional training, education, and healthcare, those who lack access to these technologies may find themselves at a significant disadvantage. This digital divide could have profound implications for employment, education, and healthcare outcomes, reinforcing existing inequalities across both social and economic dimensions. For example, students in developing regions may miss out on immersive, hands-on learning experiences, while workers in lower-income countries may lack access to advanced XR-based training programs that enhance job skills [166]. In healthcare, the inability to access XR-driven therapies or remote diagnostics could leave underserved populations behind, further widening the healthcare gap between regions [11].

To mitigate these challenges, policymakers and international organizations must prioritize infrastructure development, particularly in rapidly developing regions. Initiatives to expand high-speed internet access and make XR hardware affordable through public-private partnerships and government subsidies are critical [162]. For example, projects such as Google's Project Loon, which delivers internet access to remote regions using high-altitude balloons, demonstrate how innovative solutions can bridge the gap in connectivity [163]. By expanding digital infrastructure, these regions can more effectively integrate XR into education and healthcare, allowing users to benefit from immersive learning environments and advanced medical diagnostics regardless of geographic limitations.

In addition, digital literacy programs must be developed and widely implemented to equip individuals with the skills necessary to navigate XR environments. For developing regions, in particular, this is essential to ensure that users are not just passive consumers of XR but can also engage meaningfully with the technology to improve their education, health, and employment opportunities [166]. These programs should focus on both the technical aspects of XR use and the ethical implications of data sharing and privacy within virtual environments. As XR technologies become more integrated into daily life, ensuring equitable access and understanding of these technologies will be essential for fostering inclusive development [152].

Lastly, international collaboration is necessary to ensure that XR technologies are deployed ethically and inclusively in developing regions. Industry stakeholders, governments, and NGOs must work together to develop policies and initiatives that prioritize access to XR technologies while safeguarding user rights and privacy. This includes extending the principles of global data protection regulations, such as the GDPR, to developing countries where legal protections may be weaker, ensuring that XR platforms operate ethically and that users' privacy and autonomy are protected [161].

5.3.2. Future Integration with Artificial Intelligence (AI)

The future of XR technologies lies in their integration with Artificial Intelligence (AI). AI-driven XR environments will have the capability to adapt to users' cognitive and emotional states in real time, creating personalized experiences that optimize learning, productivity, and entertainment [167]. For instance, AI systems could detect when a user is mentally fatigued and adjust the virtual environment to provide less cognitively demanding tasks [67]. Alternatively, in professional settings, AI could assist users in solving complex problems by analyzing their cognitive patterns and providing real-time feedback [155].

However, the integration of AI into XR also raises concerns about autonomy and decision-making. As AI systems become more sophisticated, there is a risk that users may become overly reliant on these technologies for cognitive support, potentially diminishing their problem-solving abilities and critical thinking skills [167]. The question arises: how much cognitive control should we delegate to AI-driven systems, and at what point does this delegation hinder, rather than augment, human cognition?

## 6. Conclusion

As demonstrated in this review, XR has vast potential across various domains. In education and training, XR facilitates immersive learning experiences that transcend traditional teaching methods, providing real-time, interactive environments for students and professionals alike. From virtual laboratories in STEM education to medical training that simulates high-stakes surgeries, XR fosters cognitive augmentation by allowing users to interact with complex systems in ways that were previously unimaginable. In healthcare, XR is revolutionizing therapeutic interventions, including cognitive-behavioral therapies and neuropsychological assessments, which provide more ecologically valid assessments of cognitive functions, enhancing clinical diagnosis and rehabilitation efforts. Additionally, XR's potential in professional and industrial training, as well as in entertainment and cultural heritage, underscores its wide-ranging applicability.

However, the integration of XR with emerging technologies like AI and BCIs introduces significant ethical, psychological, and societal challenges. Issues such as data privacy, the psychological effects of prolonged immersion, and the potential for social inequality must be addressed as XR becomes increasingly embedded in daily life. The risk of overreliance on AI-driven cognitive augmentation further raises questions about balancing human autonomy with technological assistance.

To fully harness the potential of XR technologies while mitigating risks, it is essential to develop robust ethical frameworks. These frameworks must prioritize the protection of privacy, equitable access to technology, and the promotion of mental health. As XR continues to evolve, the balance between cognitive augmentation and ethical responsibility will be crucial in ensuring that these immersive technologies contribute to human development rather than detracting from it. By fostering responsible use and expanding access, XR can play a pivotal role in shaping a future where digital and immersive technologies enhance, rather than hinder, human flourishing across educational, healthcare, professional, and societal contexts.

**Funding:** This research received no external funding

**Conflicts of Interest:** The author declare no conflicts of interest.